# Hyperspectral imaging with Raman scattered photons: A new paradigm in Raman analysis.

K N Prajapati[1*], Anoop A Nair[1], S Ravi P Silva[2] and J Mitra[1#]

[1]School of Physics, Indian Institute of Science Education and Research, Thiruvananthapuram 695551, India
[2]Advanced Technology Institute, University of Surrey, Guildford GU2 7XH, United Kingdom

**krishnagkp12114@iisertvm.ac.in, #j.mitra@iisertvm.ac.in*

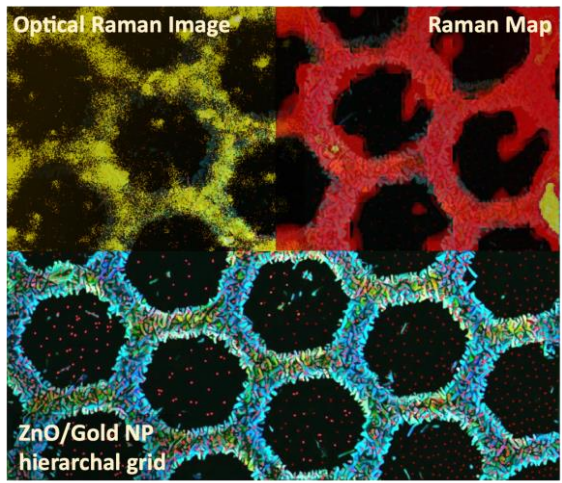

Composite image of electron microscopy image of Au nanoparticle decorated ZnO nanorods on the hexagonally patterned substrate (bottom, false colour) along with the Raman map (top right) and optical Raman image collected with Raman scattered photons (top left) with rhodamine blue coated over the substrate.

**Abstract**

Surface enhanced Raman spectroscopy, is a technique of fundamental importance to analytical science and technology where the amplified Raman spectrum of analytes is used for chemical fingerprinting. Here, we showcase an engineered hierarchical substrate in which the plasmonically active regions are restricted to a micron scale, 2D hexagonal pattern. The Raman signal enhancement of any analyte uniformly coating the substrate is consequently bears a high registry with the 2D pattern. This spatially segregated enhancement allows optical imaging of the 2D pattern solely using the Raman scattered photons from the analyte. While pattern brightness and contrast determine analyte identification and detection sensitivity, the spectrally selective contrast allows for tuning specificity. Conceptual proof of the technique is demonstrated via the acquisition of Raman images with rhodamine and fluorescein and finally applied to detect glucose in 40 mM concentration. The large area imaging and the inherent requirement of spatial uniformity for positive detection implemented using a machine learning based automated pattern recognition protocol increases the statistical confidence of analyte detection. This simultaneously multisite signal detection sacrifices continuous spectral information at the cost of speed, reproducibility and human error via automation of detection of the hyperspectral imaging technique presented here.

**Introduction**

The field of chemical sensing and analyte detection by Raman spectroscopy has been significantly advanced with the advent of plasmonics, and the discovery of surface enhanced Raman spectroscopy (SERS)[1]. Today, SERS is a powerful technique with increasing infiltration in fields such as food safety, medical diagnostics, and especially in areas requiring trace analyte detection like forensics and security. In SERS, the Raman signal enhancement originates from collective charge oscillations, i.e. plasmons, induced in metal nanostructures by electromagnetic excitation. It results in amplification and localisation of the incident electric field into subwavelength volumes within the close proximity of the nanostructures, yielding stronger Raman scattered signals from molecules lying therein. These effects have been extensively investigated, employing geometric hot-spot and antennae effects leading to unprecedented sensitivity and single molecule detection[2, 3]. Plasmonics is a platform that is enriched by allowing the extreme interplay of electrons, photons and phonons, over a large bandwidth, ensuring applications well beyond spectroscopy, from enhancing reactions to process kinetics that were hitherto deemed impractical[4, 5]. The scientific and industrial importance of the field is evidenced by the phenomenal increase in journal publications over the last decade and the ever increasing number of research spin-offs and companies dealing with SERS related products and platforms. In parallel, improvements in optical detection, including spectrometers, has led to increased adoption of the platform beyond scientific research to wider applications in sectors like agriculture, security, water contamination, biomedical applications etc. Yet, beyond the research laboratories, the technique remains unaffordable due to the high latent cost of the experimental setup and the specialised expertise and manhours involved in the subsequent analysis of the Raman spectra that lead to chemical identification.

In spite of the significant advances made in the technical aspects of SERS, there remains a veritable gap in quantitative understanding of the origin of SERS enhancement, especially the dichotomy between chemical and electromagnetic modes of signal enhancement[6-8]. Previous investigations have demonstrated that random metal nanostructure clusters often offer higher SERS enhancement compared to engineered, lithographed nanostructures, i.e. the exact shape and size of the nanostructures are of lesser importance than the statistical abundance of "hot-spots" created therein. In fabricating SERS substrates, the above observation tilts the balance away from engineered nanostructures that are ordered in the nanoscale towards disordered nanostructure clusters, which are easier to make and scalable with significantly lower fabrication costs. Further, the present investigation aims to translate the SERS technique such that it becomes more pervasive, reduces cost and involves a chemical identification protocol that is amenable to modern data analysis techniques like machine learning. Fundamentally, in SERS, the "substrate" remains the most critical component which benefits from intelligent scientific and technical design. Here we showcase a SERS substrate that is adorned with a generic $\mu$scale hexagonal metalized pattern, onto which ZnO nanorods (diameter ~ 200 nm) are electrodeposited and subsequently decorated with Au nanoparticles (NPs) of diameter ~ 30 nm, creating a hierarchical SERS substrate that is disordered in the nanoscale but ordered in the micron scale. Expectedly, the Raman signal of analytes (Rhodamine B and Fluorescein) coated onto this substrate shows large signal amplification due to the plasmonic Au NPs, enabling detection of analytes in nanomolar concentrations, as has been reported earlier[9-11]. But more importantly, the signal enhancement is restricted to the hexagonally patterned region, i.e. at the ZnO nanorods (ZNRs) decorated with Au NPs, even though the analyte is uniformly coated across the entire substrate. The enhanced Raman signal and its spatial registry with the 2D $\mu$scale pattern makes it possible to "image" the surface with Raman scattered photons through an optical microscope, thereby translating Raman spectroscopy to microscopy. The efficacy of image acquisition and analysis can be furthered by using suitable filters at selected Raman band(s) of the suspect analyte. Operationally, above threshold imaging of the characteristic pattern at multiple known wavenumber bands will trigger the positive identification of an analyte, which is demonstrated here. The quantum efficiency (QE) of Raman scattering, which is an inelastic scattering process, is notoriously low ($< 10^{-6}$). Thus obtaining a Raman image would require extreme emission amplification, well above the background light from all sources – for unlike Raman spectral analysis (peak identification), the imaging route precludes any background subtraction. We show that the hierarchical nature of the substrate and the material properties of the Si, ZnO and Au combination is such that the background does not dominate the overall emission within the imaging spectral bandwidth, which remains responsive only to the intensity of the Raman scattered photons of the analyte.

We present the motivation and methodology of the development of the 2D patterned hierarchical SERS (2D-SERS) substrate and imaging cum analysis protocol for chemical identification of analytes by hyperspectral imaging as an alternative to the standard spectral analysis. The methodology developed here is not tailored for trace chemical detection or the ultrasensitive single molecule detection, but for detection of analytes available in $\mu$M concentrations, in volumes that provide uniform coverage of the substrates developed here. The developed methodology is standardised using Raman images on known analytes and then applied to detect glucose in biologically relevant concentrations. Finally, we outline an optimised machine learning based large area pattern detection protocol that leverages the inherent advantages of the Raman imaging methodology to detect the presence of known analytes with > 90% confidence. The 2D-SERS substrates presented here are further amenable for surface functionalization to selectively bind molecules in a fluid or flow cell for the chemical identification of specific species.

**Results and Discussion**

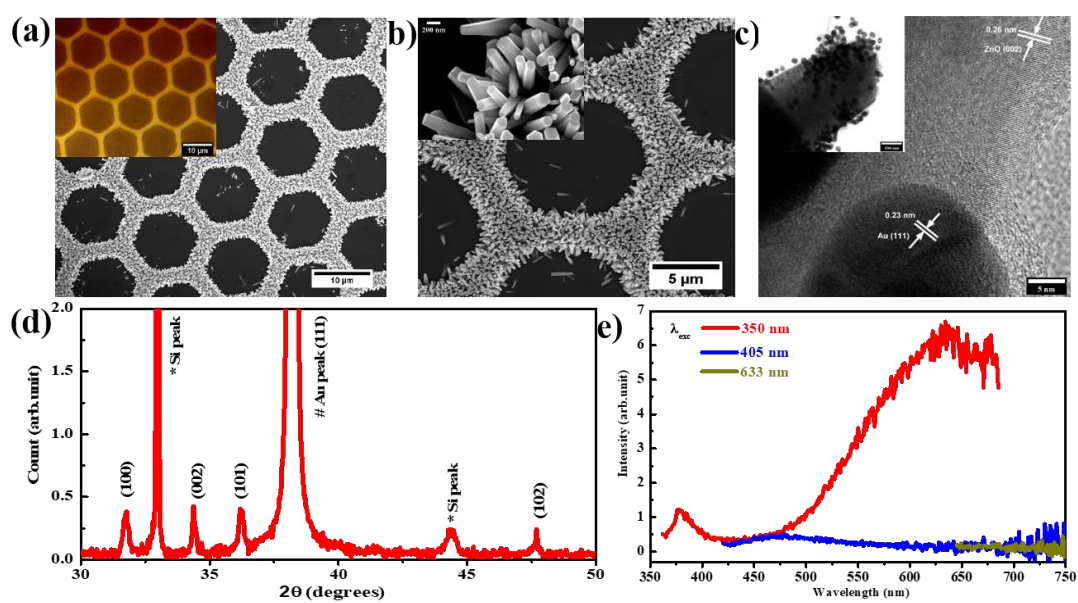

Figure 1: (a) SEM image of ZNR/Au grid/Si with an inset of optical microscope image of Au grid/Si pattern, (b) High magnification SEM image of a single hexagonal ZNR grid and Inset shows close up ZNR (scale bar is 200 nm), (c) HRTEM image of ZNR attached with Au NP with inset of low resolution TEM image of the same, (d) XRD Spectra of ZNR/Au grid/Si substrate. (e) PL spectra of ZNR/Au grid/Si substrate at different excitation wavelength.

Figure 1(a) shows the scanning electron microscope (SEM) image of ZNRs, electrodeposited[12] on pre-patterned Au hexagonal grid on $SiO_2$/Si substrate (ZNR/Au grid/Si), with an optical image of bare Au grid/Si substrate in the inset. Figure 1(b) shows a magnified SEM image of a hexagonal ZNR/Au grid/Si pattern with the inset showing a close up of the ZNRs. The ZNRs with an average diameter ~ 200 nm and length of ~ 2 μm uniformly cover the Au grid. Figure 1(c) displays the high resolution transmission electron microscopy (HRTEM) image of a ZNR decorated with Au NPs where the lattice fringes of ZNR with spacing ~ 0.26 nm confirm the orientation and crystallinity of the ZNRs and the lattice fringes of an Au NP with a spacing ~ 0.23 nm confirming the Au(111) plane[13]. A low resolution TEM of a ZNR attached with Au NPs is shown in the inset of figure 1(c). The XRD spectrum of as grown ZNRs on Au grid/Si substrate is shown in figure 1(d), where peak indexing confirms the wurtzite structure of ZNRs. The Au (111) and Si peaks arise from the underlying Au layer and the Si substrate. Figure 1(e) shows a set of photoluminescence (PL) spectra for ZNR/Au-grid/Si system for various excitation wavelengths ($\lambda_{exc}$). For the excitation energies greater than the ZnO bandgap energy ($E_{exc} > 3.3$ eV), the emission spectra show a sharp UV peak at 377 nm corresponding to the near band edge (NBE) emission and a broad visible emission centred at 645 nm attributable to oxygen vacancy related defect states[12, 14]. Importantly, the PL spectra for excitations with energies less than the bandgap, i.e. at $\lambda_{exc}$ = 405 nm and 633 nm, show negligible emission beyond 650 nm. Figure (S1) in supplementary information shows the extinction spectra of Au NPs (dispersed in deionized water) displaying the characteristic maximum around 525 nm with (FWHM ~68 nm) due to plasmonic absorption with the inset showing the TEM image of the Au NPs of average diameter ~ 30 nm.

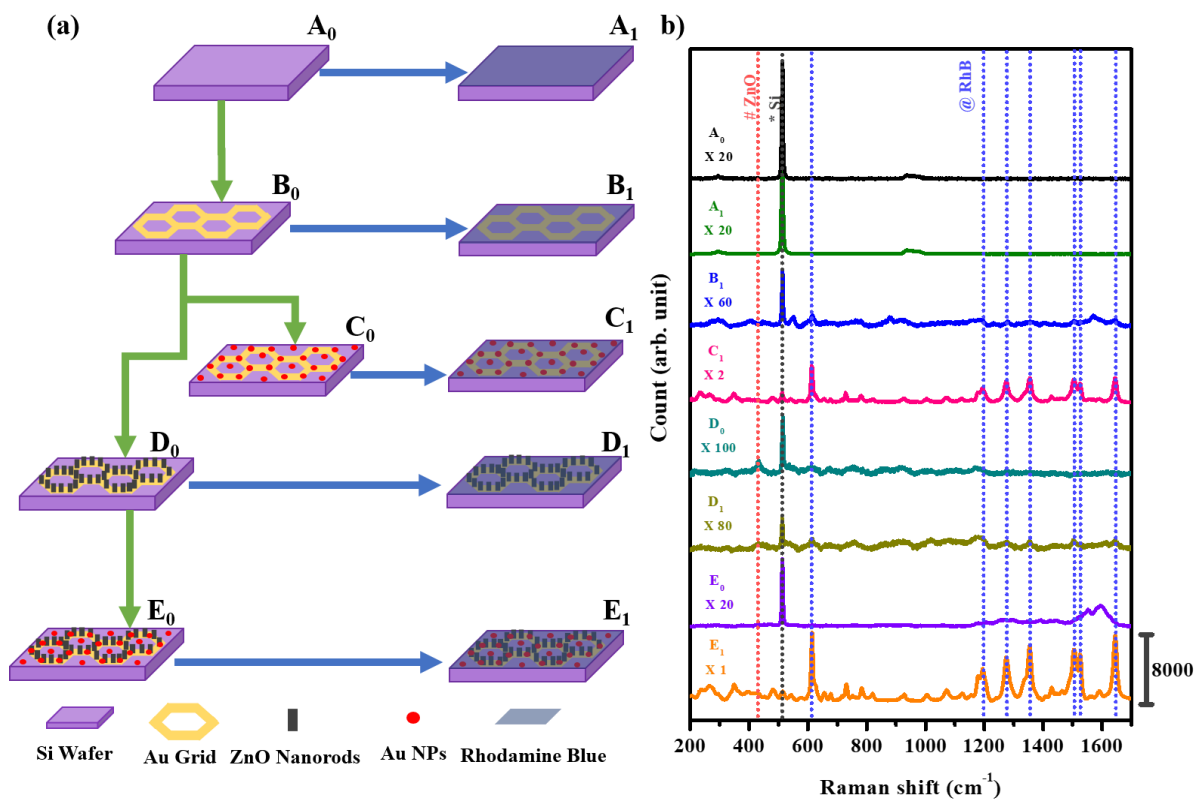

Figure 2: (a) Development flowchart of the SERS substrate from bare Si ($A_0$) to AuNP/ZNR/Au-grid/Si ($E_0$). Their analyte (Rhodamine B) coated counterparts have subscript 1. (b) Background corrected Raman spectra of selected bare and analyte coated substrates. The spectra have been vertically shifted for clarity and all spectra (barring $E_1$) have been amplified by a multiplication factor for comparison.

The schema in figure 2(a) shows the steps of development of the 2D-SERS substrate through its levels of nanostructure incorporation and hierarchy, with the subscripts 0 and 1 demarcating the bare and Rhodamine (RhB) coated samples. $A_0$: bare Si, $A_1$: RhB/Si, $B_0$: Au-grid/Si, $B_1$: RhB/Au-grid/Si, $C_0$: AuNP/Au-grid/Si, $C_1$: RhB/AuNP/Au-grid/Si, $D_0$: ZNR/Au-grid/Si, $D_1$: RhB/ZNR/Au-grid/Si and finally the 2D-SERS substrate configured as $E_0$: AuNP/ZNR/Au-grid/Si with its RhB coated counterpart $E_1$: RhB/AuNP/ZNR/Au-grid/Si. Figure 2(b) shows the background corrected Raman spectra from some of the above samples. Spectra with subscript 1 correspond to substrate coated with 200 μM aqueous solution of RhB, with all other experimental parameters remaining the same. The Si Raman peak at 514 cm$^{-1}$ dominates the acquired spectra in all cases barring that labelled $E_1$ from the final 2D-SERS substrate coated with RhB. In the latter RhB's plasmonically amplified Raman peaks dominate the overall emission spectrum. The identical nature of the $A_0$ and $A_1$ spectra reflects the low QE of the Raman scattering process such that no signature of RhB is detectable in $A_1$. The spectrum for $B_1$ (RhB/Au-grid/Si) shows the first signatures of the Raman peaks associated with RhB due to SERS enhancement from the roughened Au-grid layer. However, the signals are weak since the Au-grid covers less than 20% of the substrate area. The same patterned substrate, when decorated with Au NPs ($C_0$), shows a significantly larger SERS activity yielding the RhB spectrum labelled $C_1$. Here the plasmonic activity of Au NPs enhances the RhB Raman peaks, which now become well resolved, though the Si substrate peak still remains detectable. The Raman spectra of bare ZNR samples ($D_0$) and the RhB coated $D_1$ provides a weak signature of the strongest of the Raman active modes of ZnO, i.e. the $E_2$ peak around 440 cm$^{-1}$ correspondings to in-plane motion of oxygen atoms in ZnO$^{(15)}$. The Raman spectrum of the bare final 2D-SERS substrate ($E_0$) is again dominated by the sharp Si peak (514 cm$^{-1}$) along with a broad structure around 1500 cm$^{-1}$, which have been identified to arise from the ligands attached to the Au NPs. The spectrum from sample $E_1$ (RhB on $E_0$) fully resolves the Raman signature of RhB, displaying all the known peaks of the analyte. The spectra have been numerically scaled by the factors mentioned for comparison with that from $E_1$. Table (S1) in the supplementary information lists all the peaks and their origin, benchmarking them against known datasets.

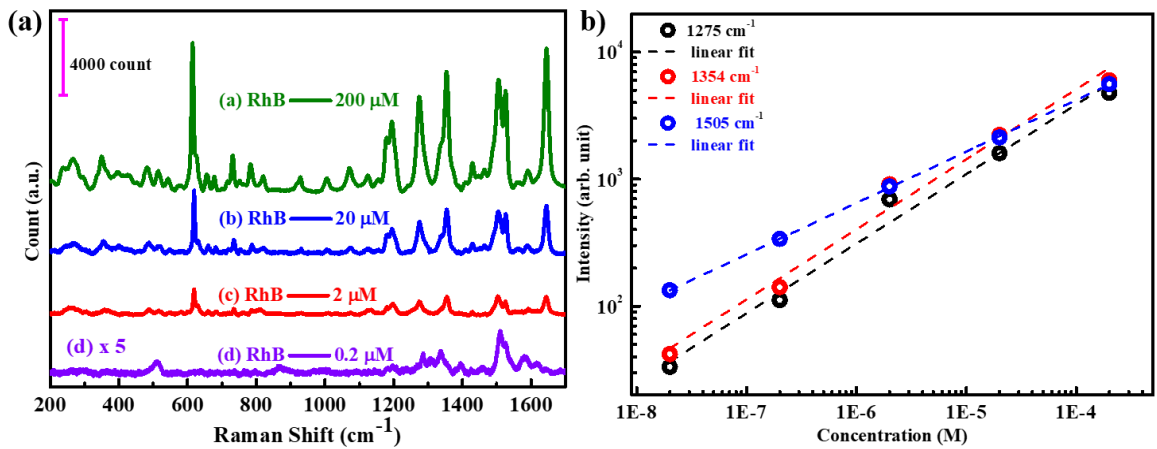

Figure 3: (a) SERS spectra of RhB with different concentrations on the substrate $E_0$. (b) The log-log plot of RhB concentration *vs.* SERS intensity at 1275 cm$^{-1}$, 1354 cm$^{-1}$ and 1505 cm$^{-1}$ wavenumbers, the dotted lines are linear fits to data points.

Figure 3(a) shows a series of SERS spectra obtained from RhB of different concentrations varying from 200 μM to 0.2 μM on the 2D-SERS substrate. Evidently, the intensity of the Raman peaks decreases with the decreasing concentration of RhB with almost no detectable change in the Raman shift (peak position). Figure (3b) shows the log-log plot of the variation of peak intensity with concentration at three wavenumbers 1275 cm$^{-1}$, 1354 cm$^{-1}$, and 1505 cm$^{-1}$. The plot follows linear dependency between the two quantities in the logarithmic scale, indicating that the peak intensity ($I$) is related to concentration ($C$) of the analyte via a mathematical relation of the form $I = \alpha C^\rho$. Where $\alpha$ and $\rho$ are peak dependent fit parameters. Such dependencies are useful as calibration parameters for quantifying unknown analyte concentration$^{(9-11)}$. The above results demonstrate the quality of the hierarchical 2D-SERS substrate, which may be further quantified by calculating the SERS enhancement factor (EF) for the substrate according to the following standard equation.

$$EF = \frac{I_{SERS} \, N_{Raman}}{I_{Raman} \, N_{SERS}} \qquad (1)$$

Where $I_{SERS}$, $I_{Raman}$ are the intensities of a SERS and the corresponding normal peak in a Raman spectrum and $N_{SERS}$ and $N_{Raman}$ are the number of analytes (i.e. RhB) molecules illuminated within the incident laser spot$^{(16),(11)}$. Comparing the above numbers corresponding to the 0.2 μM RhB peak at 614 cm$^{-1}$ on substrate $E_1$ against that obtained for 200 mM RhB on bare Si, we obtain EF ~ 10$^6$, with comparable EF calculated for the 1645 cm$^{-1}$ peak. Spatial uniformity in the Raman enhancement and between the Raman spectra recorded across the substrate is confirmed by the series of spectra presented in figure (S2) which were recorded at 3 different positions, uniformly distributed across the 10 mm x 10 mm substrate.

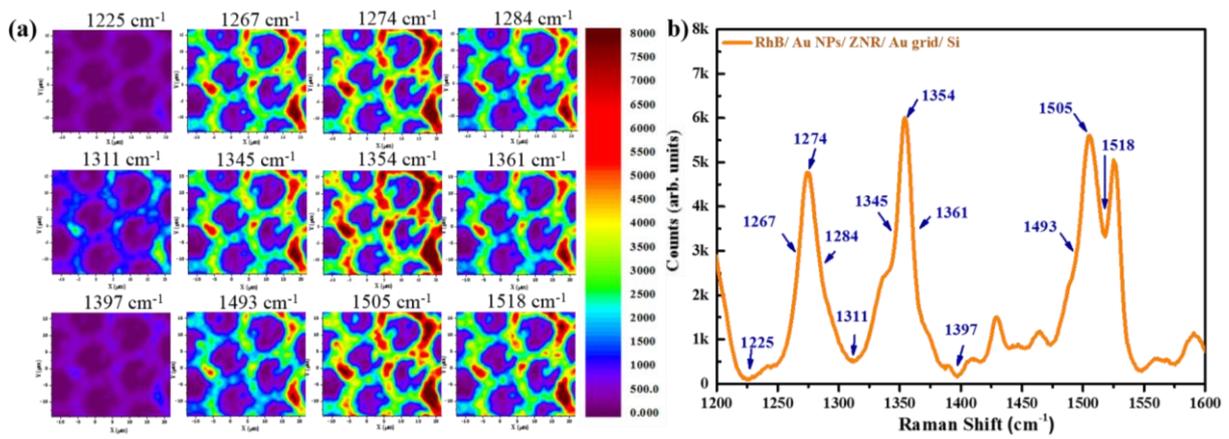

Figure 4: (a) Raman maps of RhB on Au-NP/ZNR/Au-grid/Si substrate at different wavenumbers (b) Background corrected Raman spectrum with arrows denoting the wavenumbers at which the maps were recorded.

Having established the basic performance of the 2D-SERS substrate, figure 4(a) shows a series of spatially resolved Raman intensity maps for 200 µM RhB at 12 different wavenumbers between 1200 cm$^{-1}$ and 1550 cm$^{-1}$. The maps were generated from Raman spectra acquired over a 30 µm × 30 µm area with a laser spot size of ~ 1 µm and step size of 1 µm, i.e. over 900 points, which were post-processed to yield the map. The SERS intensity distribution across the 12 maps originate from the scattered photon intensity evidenced in figure 4(b), at the highest, lowest and middle emission intensity wavenumbers across the three spectral peaks shown. It is important to note that though the spectrum in figure 4(b) is background corrected the Raman maps in figure 4(a) record the total intensity, i.e. without any background correction. Evidently, the contrast in the Raman maps shows a high degree of registry with the hexagonal pattern of the substrate that harbour the ZNRs decorated with Au NPs. The signal enhancement arises from the plasmonic Au NPs which were coated across the entire substrate, in spite of which the hierarchical nature of the substrate at the regions of ZNRs, i.e. on the hexagonal pattern, most effectively enhance and localise the Raman signal therein – which is the central tenant of this investigation and to be exploited in due course. The Raman maps at the three peak wavenumbers 1274 cm$^{-1}$, 1354 cm$^{-1}$, and 1505 cm$^{-1}$ (high intensity) spatially resolve the hexagonal patterns with reasonably uniform intensity patterns. By contrast maps at the wavenumbers with low intensity, i.e. 1225 cm$^{-1}$, 1311 cm$^{-1}$, and 1397 cm$^{-1}$, the contrast between the substrate and grid decrease significantly making the grid pattern are markedly less differentiable. Figure 4(a) also shows the Raman maps at six other wavenumbers corresponding to the middle intensities of the peak maxima at 1267 cm$^{-1}$, 1284 cm$^{-1}$, 1345 cm$^{-1}$, 1361 cm$^{-1}$, 1493 cm$^{-1}$ and 1518 cm$^{-1}$, where the contrast in the Raman maps are midway between those at the highest and lowest intensities in the spectrum, where the hexagonal grid are resolved. Figure (S3) in supplementary information shows the Raman maps and spectrum from the same sample as in figure (4) but recorded after 70 days of storage in the dark. The spectral position of the peaks remains unchanged though their maximum intensities decrease by more than a factor of 5. The reduced contrast in the Raman maps makes the hexagonal grid pattern barely identifiable but are nevertheless there. Comparison of the data presented in figures (4) and (S3) demonstrate the robustness and stability of the substrate developed here. Raman mapping for 0.2 µM RhB adsorbed on the 2D-SERS substrate along with its Raman spectrum are shown in figure (S4). Though the spectrum adequately resolves the Raman peaks, all the three maps at 1274 cm$^{-1}$, 1505 cm$^{-1}$ and 1645 cm$^{-1}$ indicate that spatial contrast between the grid and the substrate is highly compromised, in which the hexagonal patterns are difficult to distinguish.

The Raman spectrum of 200 µM RhB on the SERS substrate (figure 2) shows that the analyte's major Raman peaks typically lies between 600 cm$^{-1}$ to 1700 cm$^{-1}$, corresponding to a 100 nm wavelength window $\Delta\lambda_{Raman}$ ~ 650 – 750 nm, corresponding to 633 nm excitation. Figure 5(a-c) shows an optical image recorded with Raman scattered photons in the $\Delta\lambda_{Raman}$ spectral window, for three different laser powers. The Raman images gain clarity and contrast with increasing 633 nm laser power from 2% – 60%, with figure 5(c) fully resolving the hexagonal grid, the regions from which the highest intensity of the scattered photons originate. To confirm that the brightness and contrast in these images dominantly arise from the Raman scattered photons from RhB, the Raman image from the bare 2D-SERS substrate, i.e. without RhB, at 60% laser power was recorded as shown in figure 5(d). This image quantifies the background illumination of all the Raman images, which originate from all scattering or emission sources inherent to the substrate, under 633 nm excitation. Of particular relevance is the low luminescence of ZNRs under 633 nm excitation (figure 1(d)). Figure (S7) and (S8) shows two further sets of Raman images on freshly prepared substrates demonstrating the overall reproducibility and robustness of the observations.

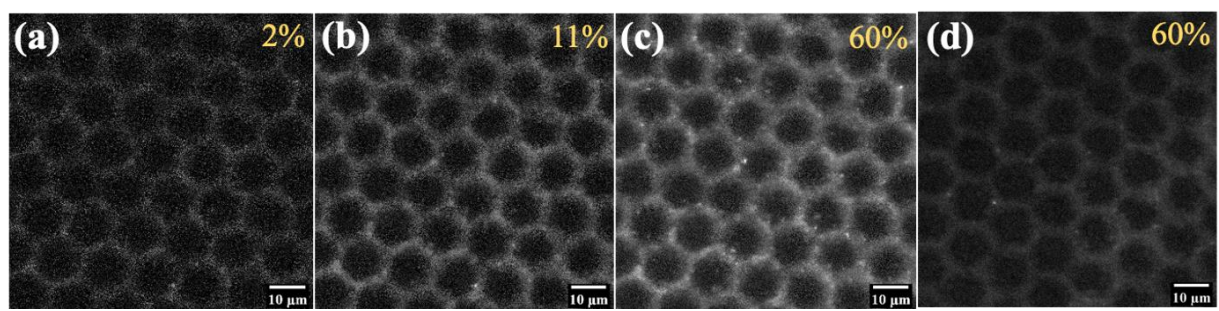

Figure 5: Optical images of Raman scattered photons of RhB. Raman images of (a-c) 200 µM RhB on patterned SERS substrate at three different 633 nm excitation laser powers. (d) shows the image for the bare substrate without any RhB. Images recorded with spectral filter 650 – 750 nm.

Figure S9 shows the Raman spectra from 200 µM fluorescein (FSN) adsorbed on the 2D-SERS substrate. Figure (6) shows the Raman images recorded with the FSN dye on the 2D-SERS substrate, again evidences the registry of the recorded pattern with the underlying hexagonal patterning of the substrate. Figures 6(c) and 6(d) were recorded at different places on the same substrate demonstrating possible variability of response across the substrate. This also demonstrates that the Raman imaging technique discussed here is applicable to multiple analytes. To further quantify the change in the brightness of the images shown in figures 5, S7, S8 and 6, a frequency distribution of pixel brightness in these images are shown in figures S10 and S11, along with the variation in cumulative brightness with laser power. The brightness range in each image is identically restricted to 0 – 100.

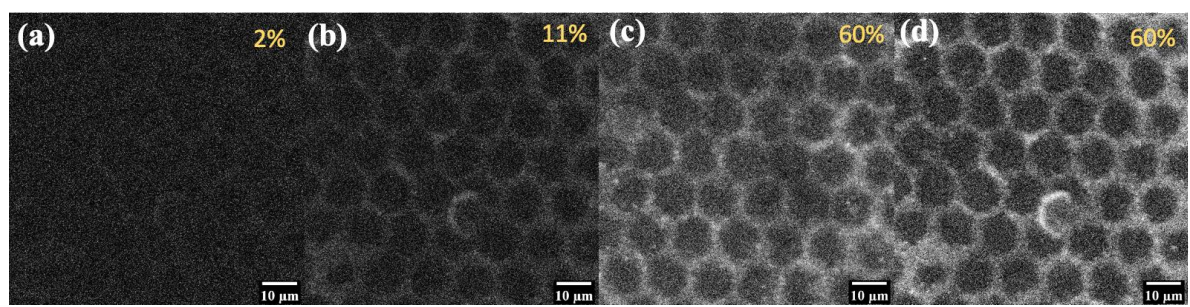

Figure 6: (a-d) Raman images of fluorescein on 2D-SERS substrate under 633 nm excitation, for three different laser power. Images in (c) and (d) were recorded at different places on the same sample. Images recorded in the range 650 – 750 nm.

It is quite likely that within the wide spectral window of detection ($\Delta\lambda_{Raman}$ ~ 100 nm), multiple analytes have different spectral peaks that contribute towards the pattern brightness and its generation. Thus mere detection of the 2D hexagonal pattern with photons detection over the large wavelength range will not result in positive analyte identification with "high confidence". To increase the confidence level of positive identification of any analyte with a known Raman signature, it will be advantageous to conduct hyperspectral imaging within multiple narrower $\Delta\lambda$. Figure 7 (a-e) further shows a series of spectrally filtered 2D-SERS images that are recorded with different spectral filters, each showing the hexagonal pattern but with different intensity corresponding to the intensity of the Raman signal in the respective spectral range of the spectrum in figure 7(f). Though these images were acquired with user selectable monochromators, they demonstrate that for identifying a specific analyte, multiple narrow bandwidth optical filters ($\Delta\lambda_{filter}$ ~ 10 nm) at designated wavelengths may be employed for positive chemical identification with high confidence. Through a microscope, the detector (human eye or camera) then looks for the hexagonal pattern in the acquired images recorded with various filters. Positive pattern recognition with above threshold brightness and contrast across multiple filters characteristic of the analyte would result in positive analyte identification.

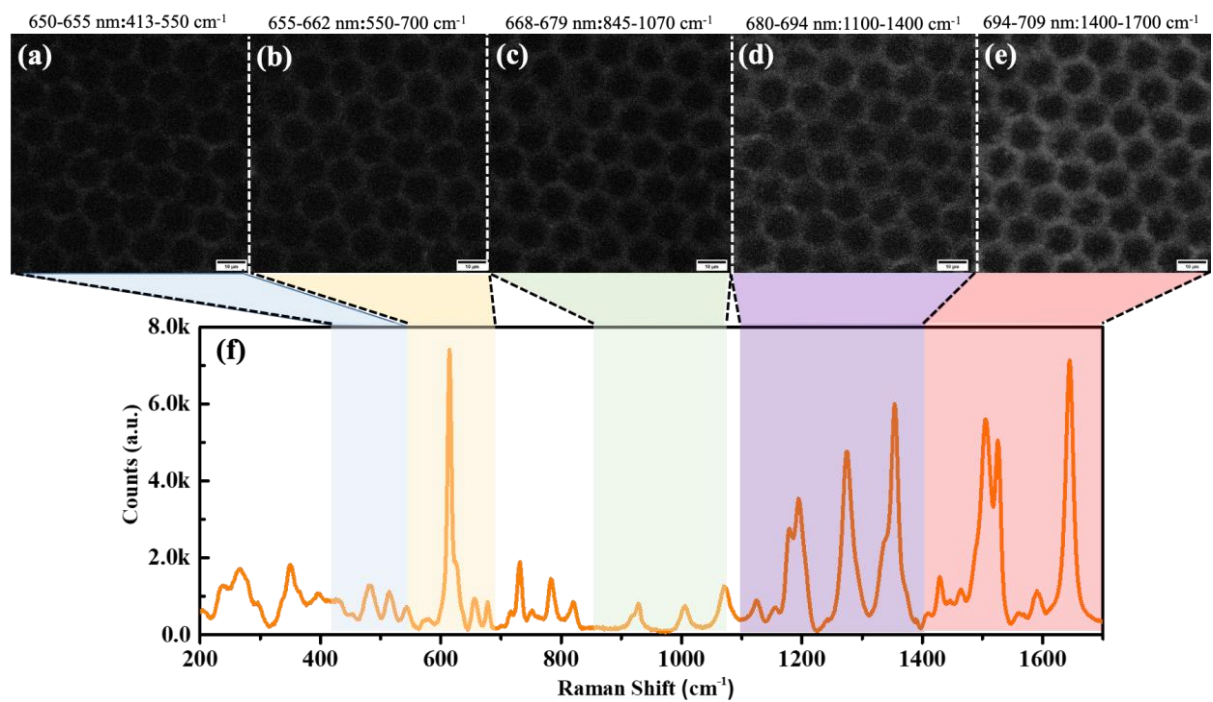

Figure 7: (a-e) Spectrally resolved Raman images of Rhodamine B on patterned SERS substrate with 633 nm excitation, corresponding to the spectral ranges demarcated in the Raman spectrum in (f). The spectral band for imaging and the corresponding Raman shift wavenumber ranges are given atop each image from (a-e).

Raman scattering based chemical identification of the kind proposed here has obvious advantages over the prevalent method of spectral analysis. Firstly, the throughput of this imaging technique is far higher than that of acquiring individual spectra at multiple positions across a sample surface for obtaining cumulative response or to generate a 2D map. For example, to generate a set of Raman maps as shown in figure 4(a), the typical time taken for data acquisition on a 30 μm × 30 μm area is 2 – 3 hrs with additional 2 – 3 hrs spent by a trained expert to process the data and generate the plots. By comparison, the time required to obtain the full set of spectrally resolved Raman images (85 μm × 85 μm area) as shown in figures 7(a-e) was typically 15 mins followed by computerised pattern recognition and analysis. Adoption of the Raman imaging technique would thus be highly advantageous for any analyte screening process. Secondly, in the present technique of Raman imaging, the chemical fingerprint of the analyte across a large area of the 2D-SERS substrate is obtained at once, with the long range uniformity parameter and averaging built into the detection protocol. That is unless the entire pattern over the whole field of optical view of the sample "uniformly" generates Raman scattered photons and generates above threshold image, registry with the micron-scale 2D pattern will not be obtained. Lastly, since positive identification of the analyte is possible by the positive correlation of multiple Raman images with a known pattern, the entire process can be readily automated utilising standard image processing and subsequent pattern recognition via 2D correlation between the images obtained and the expected pattern – paving the way to automation, further economising the entire process of chemical fingerprinting. The primary drawbacks of this technique are that prior knowledge of the Raman spectrum is necessary to define the wavelength windows for hyperspectral imaging with

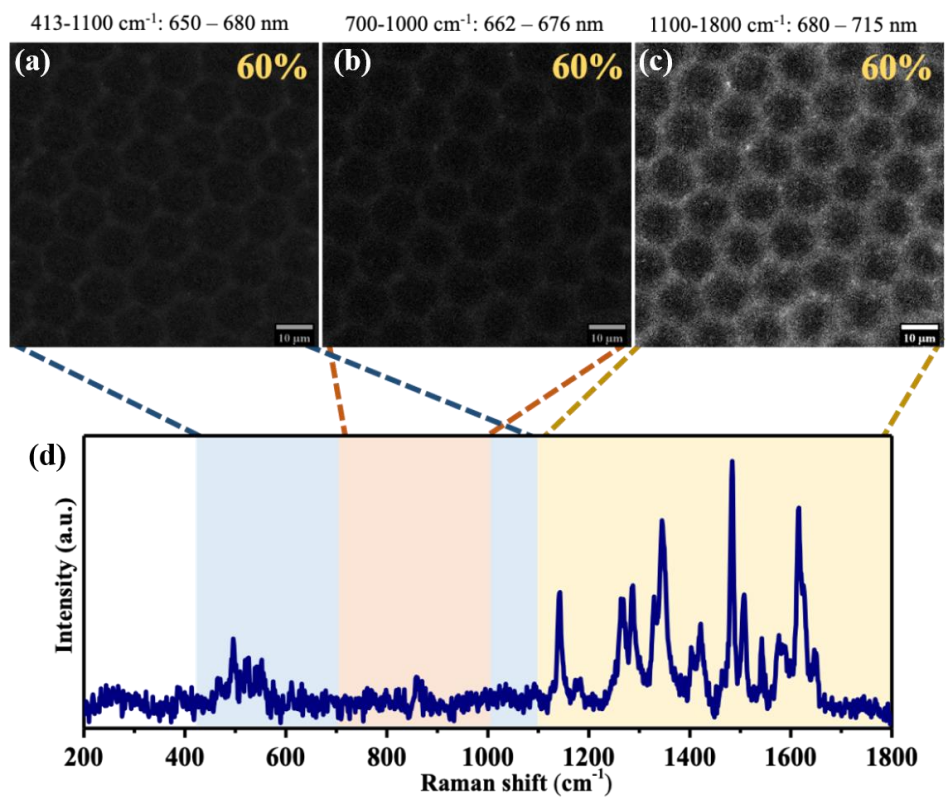

Figure 8: (a-c) Spectrally filtered images of 40 mM glucose on 2D-SERS substrate with 633 nm excitation, corresponding to the spectral ranges demarcated in the Raman spectrum in (d). The spectral band for imaging and the corresponding Raman shift wavenumber ranges are given atop each image from (a-c).

Raman scattered photons and the availability of suitable optical filters or monochromator. Thus, it can only be employed to detect known analytes and is of limited use in identifying unknown species. Further, the imaging protocol is conceptualised around processing and recording large area images of a periodic pattern precluding detection of analytes in trace quantitates or in low concentrations. It is also worth noting that the combination of AuNP:ZnO system in the hierarchical scaffold was chosen to minimise any background luminescence from the substrate in the 600 – 800 nm wavelength range. Luminescence from ZnO in the above range is effectively minimised by reducing the density of defect states and has negligible photoluminescence in the above window under 633 nm excitation and the plasmonic Au NPs further hinders any far field contribution[17].

Finally, to test the proposed detection technique on a diagnostically relevant analyte, we investigated the detection of dextrose (d-glucose) in deionised water. Non-invasive[18-20] and easy detection of glucose in humans has remained a challenge, with several investigations suggesting electrical[21], optical[19] and SERS[22] based spectroscopic detection. In healthy individuals, glucose is present in various body fluids in varying concentrations, e.g. blood (4.9 – 6.9 mM), urine (2.8 – 5.6 mM) and sweat or saliva (50 – 500 $\mu$M)[20, 23]. Thus detection of abnormal levels of glucose in the range 500 $\mu$M – 50 mM will be highly beneficial for monitoring type II diabetes. The typical Raman signal of glucose is inherently compromised due to the low scattering cross section of the process. Though SERS based detection amplifies the signal, it is disadvantaged due to low adhesion between glucose and Au or Ag nanostructures. Figure S12(a) shows the Raman spectra acquired from various concentrations of glucose solution in deionised water. Up to 40 mM concentration, the spectra are dominated by the features of deionised water, and the characteristic signatures of glucose are manifest only for concentrations in excess of 1000 mM. The detected peaks match published data[24] confirming the presence of D-glucose. Figure S12 (b) shows the Raman spectra from 4 mM and 40 mM glucose solutions drop-casted and dried onto the 2D-SERS substrate. Importantly, the spectral features are quite different between those obtained from glucose in solution and the solution dried onto the 2D-SERS substrate. Comparison between the spectra from 2D-SERS substrates between 4 mM and 40 mM glucose solutions show differences in spectral peak positions. Though the dominant peaks in each of the two spectra match or lie close to those reported[24-26], the relative strength of the peaks varies significantly. However, both spectra show that the majority of the dominant peaks lie in the range 1100 – 1800 cm$^{-1}$. Figure 8(a-c) shows hyperspectral images of the 2D-SERS substrate, coated with 40 mM glucose solution (after drying), recorded with 633 nm excitation. Their correspondence to the spectral range of detection of each image is indicated in the Raman spectrum shown in figure 8(d). Evidently, the image recorded in the 680 – 715 nm spectral window (figure 8c) shows the highest brightness, though the hexagonal pattern is present in all the images.

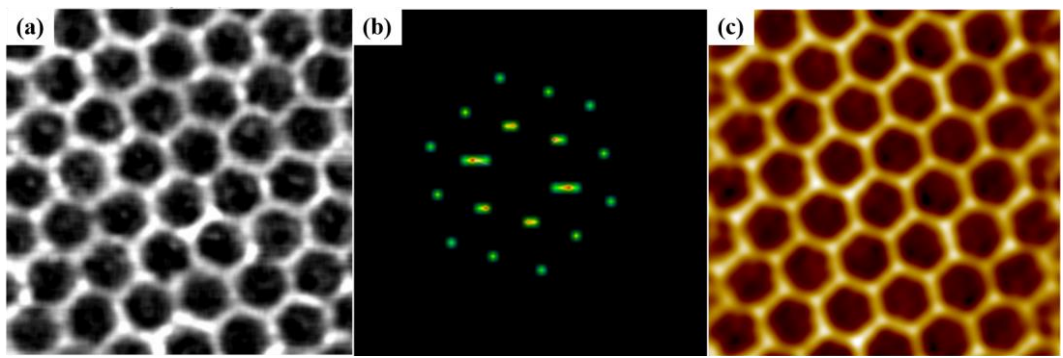

Figure 9: (a) 2D-SERS microscopy image for 40 mM glucose 2D-SERS substrate acquired with Raman scattered photons. The corresponding hcp unit cell is shown in red. (b) 2D FFT of image in (a) showing reciprocal lattice points corresponding to a hexagonal 2D pattern. (c) Inverse FFT of image (b) showing reconstructed hexagonal pattern.

Overall, the above results demonstrate that hyperspectral imaging of analyte coated 2D-SERS substrate via Raman scattered photons as presented above poses a viable route to analyte detection by automated identification of the 2D grid pattern. For automated identification of the 2D pattern, we have explored two protocols that allow quantifiable pattern recognition. Figure 9(a) shows a filtered (contrast enhanced) 2D-SERS image recorded for 40 mM glucose solution, over the spectral range 650 – 715 nm. The hexagonal closed packed (hcp) unit cell is demarcated in red. The 2D fast Fourier transform (FFT) of the image, after suitable thresholding is shown in figure 9(b), indicating that the Fourier space image of the hexagonal grid is another hexagonal lattice with lattice constant given by $4\pi/\sqrt{3}a$ and rotated through 30° about the perpendicular to the 2D real space plane, where $a$ is the lattice constant of the real space lattice (side of the hcp unit cell in figure 9(a)). Figure 9(c) then shows the inverse 2D FFT of figure 9(b), reproducing the hexagonal pattern of the original image that was acquired with Raman scattered photons. Thus, 2D FFT protocol allows adequate filtering and enhancement of the acquired image, which followed by pattern recognition, offers a straightforward and controllable option of automated detection of the 2D-SERS pattern imaged by Raman microscopy.

A second, more efficient and topical protocol in automated pattern recognition was explored utilizing a deep learning-based method, trained to identify the hexagonal pattern in the Raman images via a sequence of detection, clustering, and analysis algorithms. Across the various stages, the protocol was optimised by training on approximately 500 actual images, both with and without analytes constituting the positive and negative training sets and later tested on a new set of 150 positive and negative images for determining the accuracy of identification (as detailed in supplementary information). The Haar Cascade[27] trained on the images of interest was employed for initial detection of individual hexagons (Figure 10a), followed by clustering and analysis using the K-means clustering[28] and Convolutional Neural Nets (CNN)[29] (Figure S13 shows the workflow). Overall, the percentage of true positives (PTP) detected by the Haar cascade ranged from 55% – 100% (Figure 10a, S16(a-c)), with the percentage of false positives (PFP) detected being less than 10% (Figure S16(b)) in a majority of the images. The PFP exceeded 40% in few images (Figure S16(c)), typically associated with regions of high contrast noise (Figure S16(j)), substrate deformities and the presence of hexagonal fragments (Figure S16(i)). The problem of high PFP in such images was overcome by utilizing an intermediate shallow neural network (Shallow-CNN) trained to reject fragments and noisy patches selectively, with an accuracy of 95% (Figure S16(k)). Since the Shallow CNN was trained on an unbiased dataset, the accuracy was considered to be the sole indicator of performance. The incidence of false positives in negative images is further minimised by utilising the clustering procedure. It required the identification of cohesive groups of hexagons with a minimum cluster size of 5, with smaller clusters being rejected from further analysis, as shown in Figure 10(b). Positive images that surpassed all the above checks were assigned a confidence value using a Deep Convolutional Neural Network (Deep-CNN) [see section S2.5.2 in supplementary information]. Figure 10(c) plots the evolution of accuracy with epochs demonstrating the stability of the detection protocol. Figure 10(c) inset shows the values of the confusion matrices corresponding to the validation and test set images, showing overall detection accuracies of 94.6% and 99.3%, respectively. The accuracy values were calculated using equation (S16) from the True Positive (TP), True Negative (TN), False Positive (FP), and False Negative (FN) values of the confusion matrices. The associated F1 scores of 0.9487 and 0.9934, close to 1, for the validation and test sets (Table S4) further validates the Deep-CNN's optimal performance. It is worth noting that the algorithm is not specific to detecting hexagonal structures and is readily retrained on images with other periodic structures and symmetries, albeit of aspect ratio ~ 1. However, it is also well recognized that deep learning based pattern recognition protocols are easy to be misguided by generic noise and random areal contrast, fool-proofing which remains unexplored in the above scheme. It is anticipated that integration of the Fourier filtering scheme outlined earlier can further increase the overall accuracy of detection. Lastly, we have not attempted to discriminate between multiple analytes, which would be the next significant step change towards translating the above to applications.

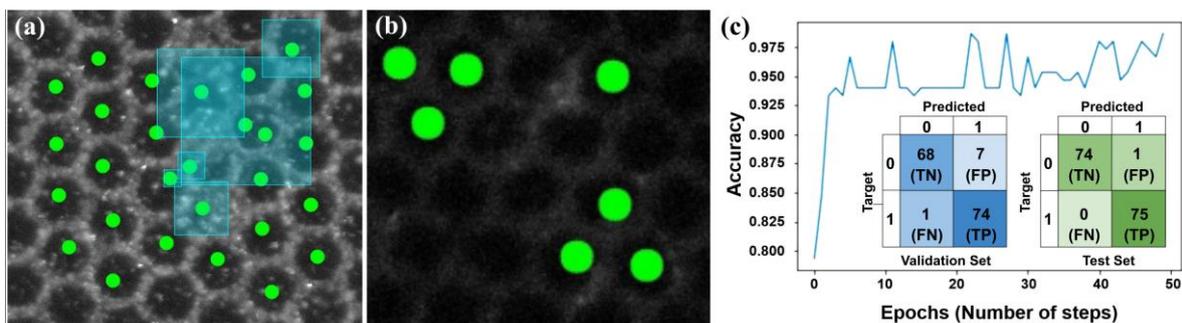

Figure 10: Automated detection of individual hexagons (green dots) by the Haar Cascade operating on (a) a positive Raman image, 22 correct and 6 wrong detections (transparent blue box) PTP = 78.5% and PFP = 21.4%, (b) a negative Raman image showing 3 clusters with less than 5 hexagons each. (c) Accuracy-Epoch plot and the confusion matrices of the Deep-CNN for the validation and test images.

**Conclusion**
This study presents a novel methodology in translating Raman scattering based analyte detection from spectroscopy to microscopy. A hierarchical substrate, patterned in the 2D on the micron-scale, is developed to showcase the methodology, central to which is the plasmonic enhancement of the Raman signal, spatially localised to the 2D pattern. Later, an imaging cum analysis protocol based on hyperspectral imaging of the substrate under electromagnetic excitation is shown to yield optical images of the 2D pattern, acquired solely with Raman scattered photons. The 2D-SERS substrate is also shown to detect analytes down to few nM concentrations with spectral signal enhancement ~ $10^6$, in the spectral range corresponding to 633 nm excitation. While the specific combination of the materials employed in the development of the 2D-SERS substrate viz. Si, ZnO nanorods and Au nanoparticles ensure minimal background luminescence within the detection window investigated here. The results do not limit the scope of exploring other materials and geometries in replicating the concept. Indeed, 2D-SERS substrates prepared employing specific material nanostructures would make them suitable for surface functionalization to selectively bind molecules in a fluid or flow cell for chemical identification of specific species and across different spectral windows corresponding to specific excitation wavelengths.

Here, the performance of the developed methodology is established through recording Raman spectra, maps and finally optical images of the substrate coated with two dyes, rhodamine and fluorescein and then applied to detect glucose as the analyte in biologically relevant concentrations (40 mM and 4 mM). Hyperspectral imaging of the substrate using narrow band filters (10 – 15 nm) corresponding to the Raman bands of the analyte under investigation is demonstrated, showing that the methodology is customisable to specific analytes. Observation of the hexagonal pattern across multiple windows then increases the confidence of analyte detection and also ensures specificity, though the later has not been investigated in detail. In all cases, the spectrally filtered images record the spatial localisation of Raman scattering to the hierarchical grid, demonstrating the conceptual basis of the presented methodology. Finally, an automated pattern recognition protocol is developed via a workflow using cascade, clustering and neural net algorithms to identify and analyze the quality of hexagonal objects in optical microscope images. The protocol trained and tested on a large set of positive and negative images demonstrates positive detection capability with 95% accuracy. It is anticipated that this investigation will help translate the SERS technique such that it becomes more pervasive, involving lower instrument costs wherein the chemical identification protocol is dominantly machine recognizable via image processing.

**Experimental Details**

**Materials:** All chemicals including zinc nitrate hexahydrate, hexamethylene-tetramine (HMTA), Au nanoparticles (NPs) (#741973), Rhodamine B (RhB, $C_{28}H_{31}ClN_2O_3$) (#83689), Fluorescein (FSN, $C_{20}H_{12}O_5$) (#46955) and D-(+)-Glucose (#G8270) of purity > 99.5% were supplied by Sigma-Aldrich and used as supplied. Deionized water was used throughout the experiments.

**Fabrication of SERS substrates:** Si substrates were patterned with a micron-scale hexagonal patterns and coated with 50 nm Au film (optical image of a pattern after mask lift-off shown in figure 1a) that acted as an electrode for electrodeposition of ZNRs selectively at the metalized region. The ZNRs were synthesized in an electrochemical bath of 5 mM equimolar aqueous solution of zinc nitrate hexahydrate and HMTA at 85 $^0$C, using platinum wire as the counter and reference electrode. The patterned substrate was sequentially biased at -2.73 V and -2.15 V for the 1 min and 30 mins, respectively[12]. After ZNR growth, the substrate was rinsed in deionized water and dried at 65 $^0$C for 1 hour. The SEM images in figures 1b and 1c show selective growth of randomly oriented ZNRs at the hexagonal patterned electrodes. Finally, to prepare the plasmonically active substrates, 30 μl of Au NPs suspension was drop-casted on top of ZNRs grown substrate to realise the hierarchical SERS substrates AuNP/ZNR/Au grid/Si. In substrates were developed on 10 mm x 10 mm sized Si pieces. For conducting the Raman studies, the various concentration of the dyes, Rhodamine B (RhB) and Fluorescein (FSN) were drop casted on the various test and control substrates. Both RhB and FSN were chosen for characterising the substrates developed since (i) both dyes do not absorb any light at the primary Raman excitation wavelength 633 nm employed here and consequently does not have any luminescence in their respective Stokes shifted Raman emission bandwidth and (ii) they have multiple Raman active modes within the Raman shift wavenumbers 600 – 1600 cm$^{-1}$ which corresponds to the wavelength range 650 – 715 nm for 633 nm excitation.

**Characterization**: Morphological imaging of the substrates and their material components was performed using Nova Nano SEM 450 field-emission scanning electron microscope (SEM) and 300 kV FEI TECHNAI G2-TF-30 transmission electron microscope (TEM). X-ray diffraction (XRD) analysis was carried out using a powder X-ray diffractometer (Empyrean, PANalytical) with reference radiation of Cu Kα = 1.540 Å at an operating voltage of 45 kV. Photoluminescence (PL) spectra were recorded using a spectrofluorometer (Fluorolog 3, Horiba Jobin-Yvon) at room temperature. The various absorption spectra were obtained using Perkin Elmer Lambda 900 spectrophotometer. All the Raman spectra and Raman maps were acquired using Horiba Scientific Xplora Plus with a 100x objective with 633 nm laser, at 1% of the total power (18 mW), spot size ~ 1 μm, typical acquisition time ~ 2s, grating - 1800 grooves/mm, unless otherwise specified. The Raman maps were recorded over approximately 30 μm × 30 μm area with a lateral step size of 1 μm. The optical Raman images were recorded using a confocal microscope (Carl Zeiss LSM 880) with 405 nm (5 mW) and 633 nm (5 mW) laser sources as specified with a wide pinhole of ~ 60 μm. Pattern recognition: The various algorithms for detecting and validating the hexagonal clusters were trained and executed on a quad-core i3-5005U CPU and by utilizing online resources like Google Colab[30].

**Author Contributions**
The project was conceptualised by JM with the methodology developed by SRPS and JM. KNP conducted the sample preparation and experimental data acquisition. AAN developed the machine learning protocol for automated pattern recognition. Data analysis performed by KNP, AAN and JM. The manuscript was written by all authors.

**Acknowledgements**
The authors thank Jervis Leo Fernandes from IISER Thiruvananthapuram for help in the acquisition of the optical images. The authors acknowledge financial support from the Royal Academy of Engineering, Newton Bhabha Fund, UK (IAPPI_77) for enabling international exchange. JM acknowledges financial support from SERB, Govt. of India (CRG/2019/004965).

# Supplementary information

# Hyperspectral imaging with Raman scattered photons: A new paradigm in Raman analysis.

K N Prajapati[1*], Anoop A Nair[1], S Ravi P Silva[2], and J Mitra[1#]

Email: * krishnagkp12114@iisertvm.ac.in, # j.mitra@iisertvm.ac.in

[1]School of Physics, Indian Institute of Science Education and Research, Thiruvananthapuram 695551, India
[2]Advanced Technology Institute, University of Surrey, Guildford GU2 7XH, United Kingdom

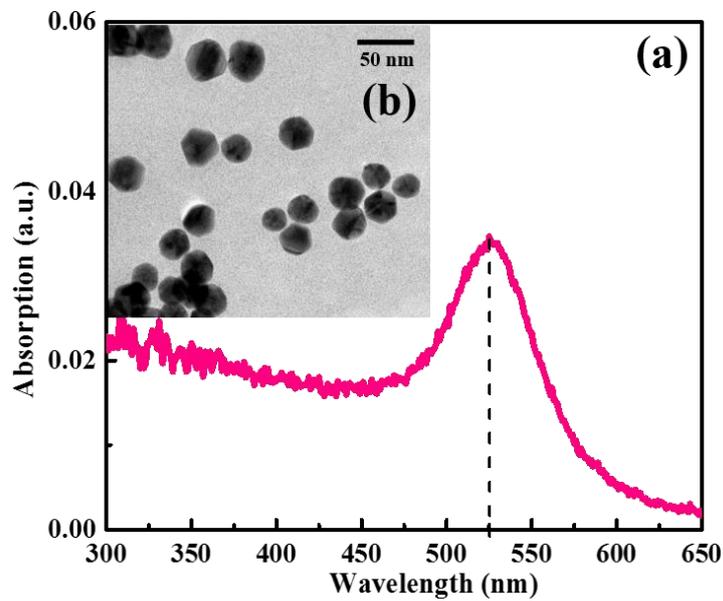

Figure S1: (a) Absorption spectrum of Au nanoparticles in DI water, (b) Inset shows the TEM image of the nanoparticles with diameter ~ 30 nm.

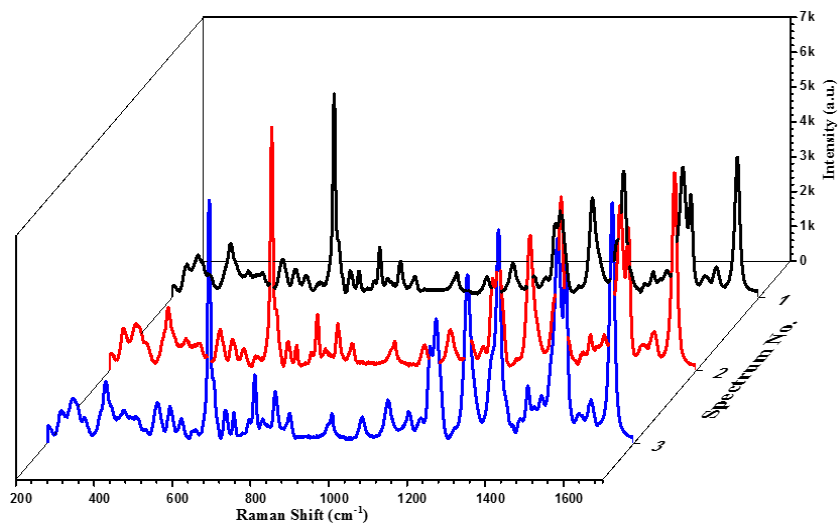

Figure S2: Raman spectra of Rhodamine B recorded on sample $E_1$ at three different places on the 10 mm x 10 mm substrate.

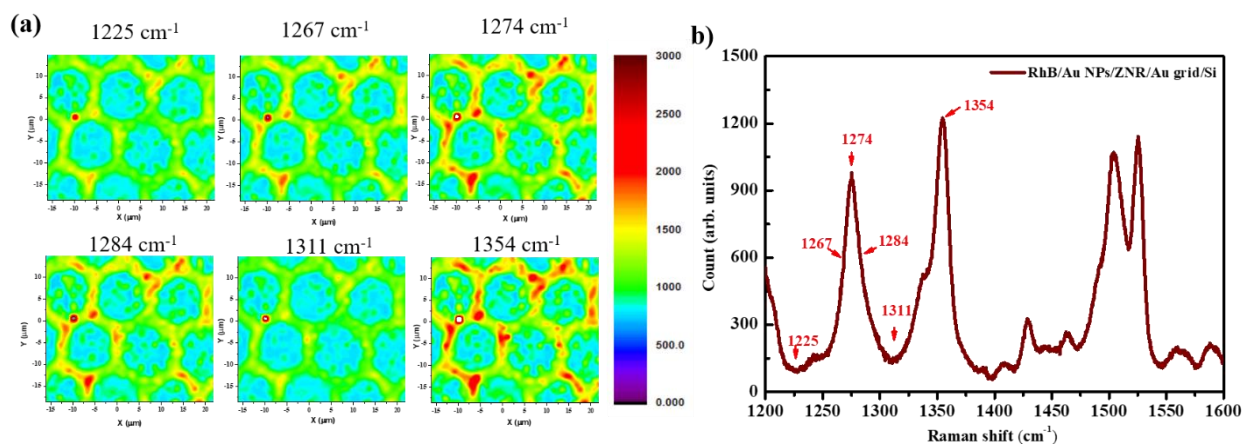

Figure S3: Data corresponding to the same sample in figure 4 but recorded 70 days later, other parameters remaining constant (a) Raman maps of RhB on AuNP/ZNR/Au-grid/Si substrate at different wavenumbers (b) Background corrected Raman spectrum with arrows denoting the wavenumbers at which the maps were recorded. These data were collected on was collected but after 70 days showing the decay in the emission intensity in the spectrum and consequential loss of contrast in the Raman maps.

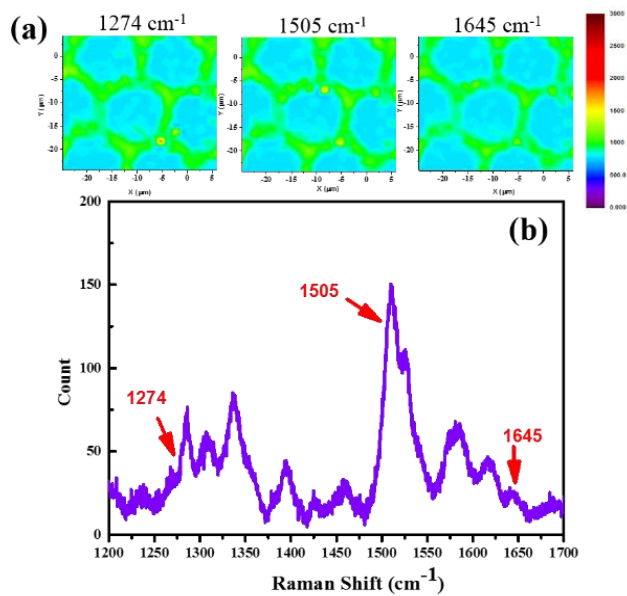

Figure S4: (a) Raman map for 0.2 µM concentration RhB (b) corresponding Raman spectrum. Arrows denote the wavenumbers at which the maps are generated.

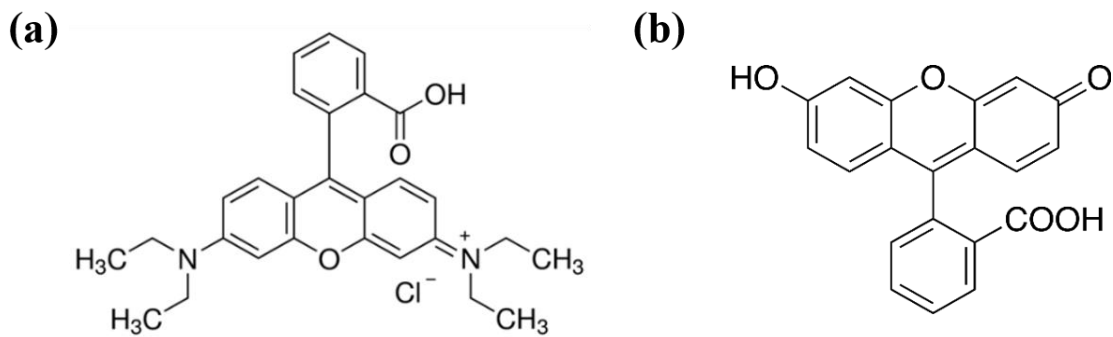

Figure S5: Chemical structure (a) Rhodamine B, (b) Fluorescein (Image courtesy: wikipedia)

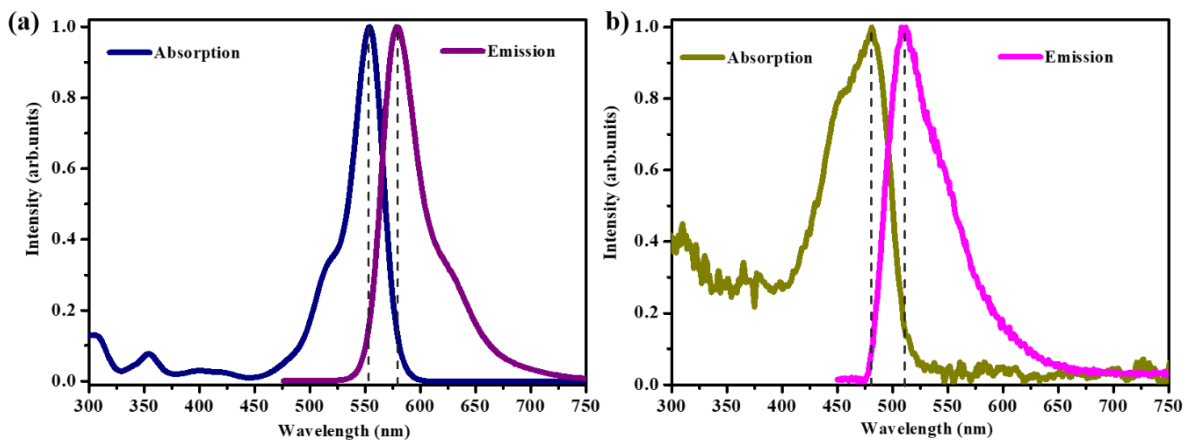

Figure S6: Absorption and Emission Spectra (a) Rhodamine B, (b) Fluorescein

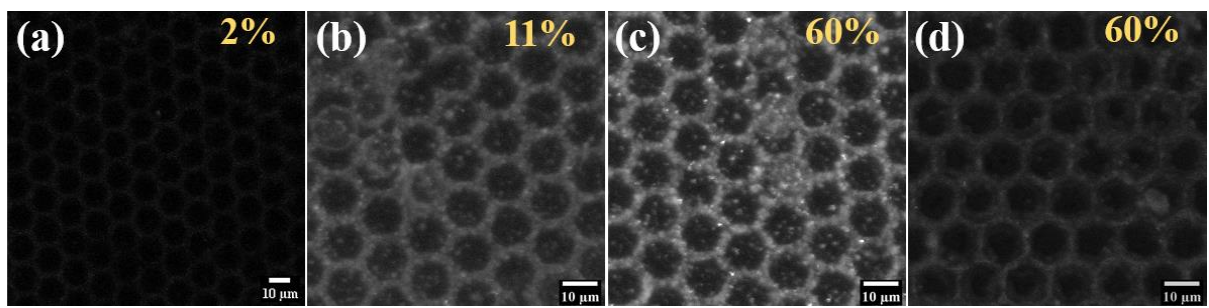

Figure S7: Optical images of Raman scattered photons of RhB. Raman images of (a-c) 200 µM RhB on patterned SERS substrate at three different 633 nm excitation laser powers. (d) shows the image for the bare substrate without any RhB. Images recorded with spectral filter 650 – 750 nm.

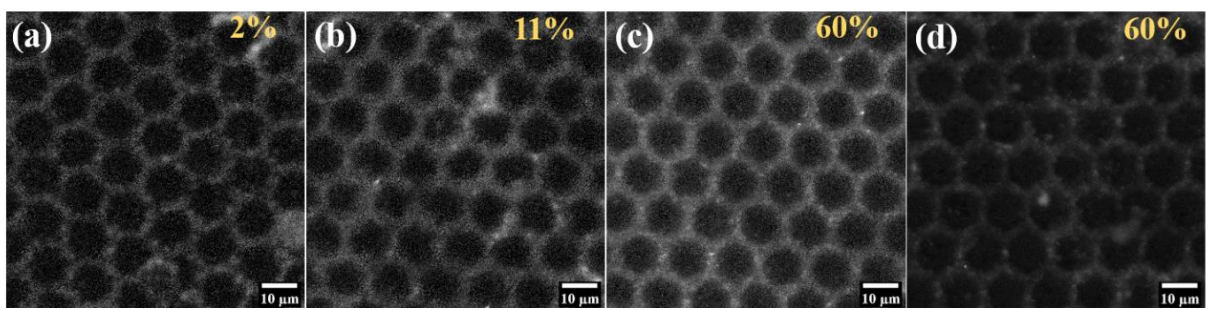

Figure S8: Another set of optical images of Raman scattered photons of RhB. Raman images of (a-c) 200 µM RhB on patterned SERS substrate at three different 633 nm excitation laser powers. (d) shows the image for the bare substrate without any RhB. Images recorded with spectral filter 650 – 750 nm.

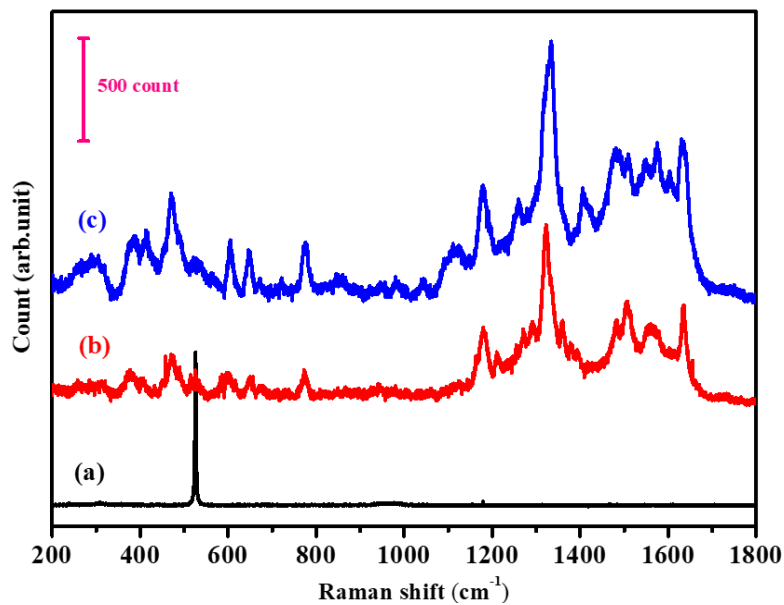

Figure S9: SERS spectra, (a) FSN/Si substrate, (b) FSN/Au NPs (30 µl)/ZNR/Au grid/Si substrate, (c) FSN/Au NPs (60 µl)/ZNR/Au grid/Si substrate

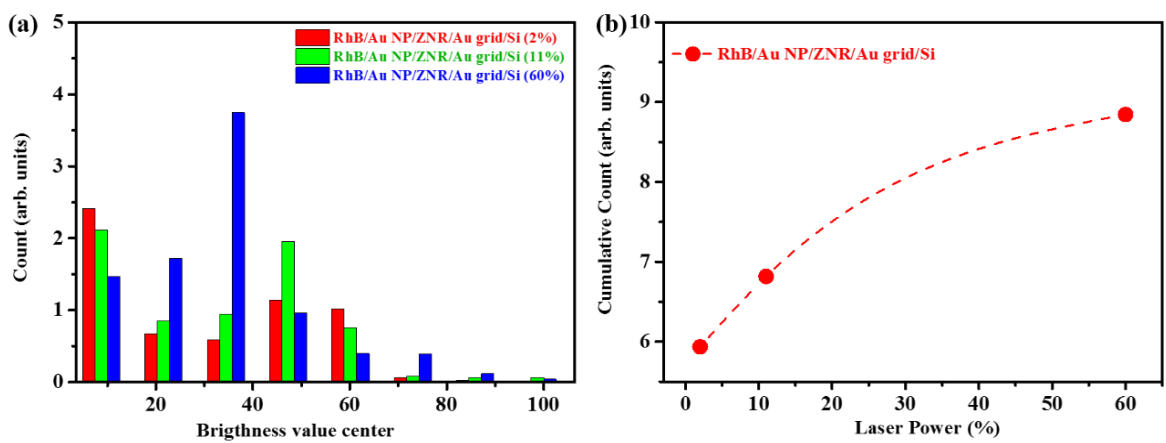

Figure S10: (a) Frequency distribution of pixel brightness level in images shown in figures 5, S7 and S8 for 2%, 11% and 60% of maximum laser power. Maximum and minimum brightness level in each image are 0 and 100 respectively. (b) Cumulative brightness counts versus laser power across the various samples discussed.

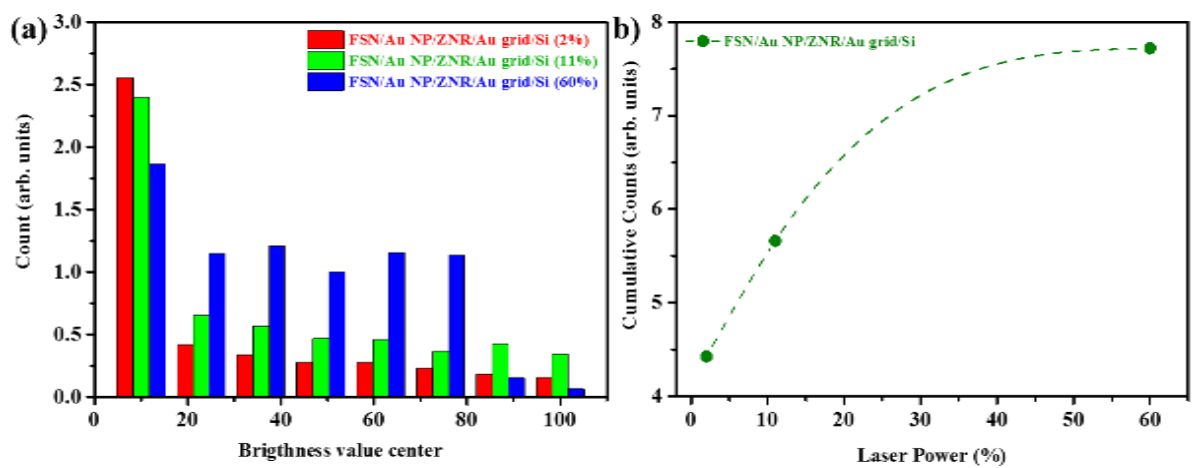

Figure S11: (a) Frequency distribution of pixel brightness level in images shown in figure 6 for 2%, 11% and 60% of maximum laser power. Maximum and minimum brightness level in each image are 0 and 100 respectively. (b) Cumulative brightness counts versus laser power across the various samples discussed.

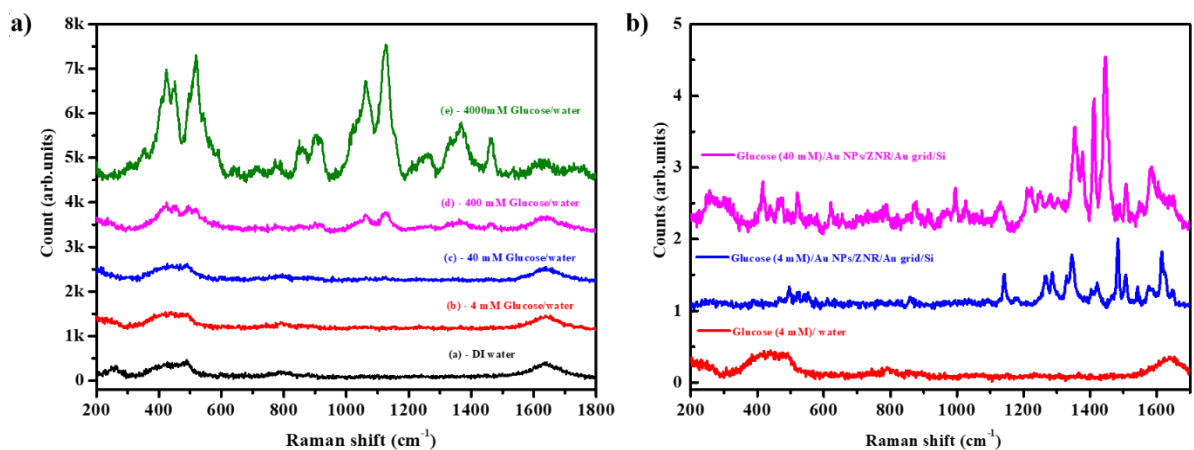

Figure S12: (a) Raman spectra of various concentrations (4 – 4000 mM) of glucose in water and bare deionised water. (b) Raman spectra of 4 mM and 40 mM glucose/water, dried on the 2D-SERS substrate

| Present work (cm⁻¹) | Literature (cm⁻¹)[1-3] | Assignments |
|---|---|---|
| 1645 | 1648 | Aromatic C-C bending and C=C stretching |
| 1590 | 1590 | C-H stretching |
| 1561 | 1564 | Aromatic C-C stretching |
| 1526 | 1530 | C-H stretching |
| 1505 | 1507 | Aromatic C-C stretching |
| 1429 | 1432 | *Peak identified but no nomenclature yet* |
| 1354 | 1354 | Aromatic C-C bending |
| 1275 | 1275 | C–C bridge band stretching and aromatic C–H bending |
| 1195 | 1193 | C–C bridge band stretching and aromatic C–H bending |
| 929 | 933 | C-H stretching |
| 782 | 785 | C-H stretching |
| 614 | 619 | Xanthene ring puckering |

Table S1: Peak positions (in cm⁻¹) in Raman spectra of Rhodamine blue from literature and present work.

| Present work (cm⁻¹) | Literature (cm⁻¹)[4] |
|---|---|
| 1632 | 1632 |
| 1602 | 1599 |
| 1575 |  |
| 1545 | 1543 |
| 1509 |  |
| 1483 | 1483 |
| 1410 | 1414 |
| 1360 | 1372 |
| 1322 | 1318 |
| 1291 |  |
| 1181 | 1185,1121 |
| 775 | 766 |
| 650 |  |
| 604 |  |
| 472 |  |

.

Table S2: Peak positions (in cm⁻¹) in Raman spectra of Fluorescein from literature and in present work

| Literature-1[5] | Literature-2[6] | Literature-3[7] | http://bme240.eng.uci.edu/students/06s/eclin/articles/StatusofGlucotion.pdf | Present Work | | |
| --- | --- | --- | --- | --- | --- | --- |
| | | | | 4M in water | 4mM on 2D-SERS | 40mM on 2D-SERS |
| 1625 | | | | 1631 | 1648 (1615) | 1648 (1620) |
| | | | | | 1578 (1507) | 1580 (1510) |
| 1455 | | 1458 | 1456 | 1463 | 1483 1421 1403 | 1455 1430 1411 |
| 1360 | 1366 | 1343 | 1365 | 1366 | 1344 | 1378 1361 |
| 1327 | | 1331 | | 1332 | 1330 | 1332 |
| 1258 | | 1272 | | | 1286 1265 | 1277 |
| 1120 | 1125 | 1119 | 1126 | 1125 | 1141 | 1128 |
| 1060 | 1060 | 1073 | 1065 | 1062 | | |
| 1035 | | 1053 (1020) | | 1022 | | 1026 |
| 910 | | 913 | 911 | 908 | | |
| 893 | | | | | | |
| 860 | | | 854 | 850 | 859 | |
| 842 | | 841 | | | | |
| 770 | 796 | 770 | | | | |
| 710 | | | | | | |
| 645 | | 653 | | | | 620 |
| 585 | | | | | | |
| 540 | | 539 | | | | |
| 518 | | | 525 | 519 | | 519 |
| 450 | | 439 | 456 (436) | 450 | 495 | |
| 424 | | 422 (404) (396) | | 423 | | 416 |
| 355 | | | | | | |
| | | 288 232 | | | | |

Table S3: Peak positions (in cm$^{-1}$) in Raman spectra of Glucose in DI water from literature and in present work. Data in the last two columns are recorded with analyte on 2D-SERS substrates.

# Deep Learning based Automated Pattern Recognition

## S1. Introduction

The detection of objects in images has been of interest for many domains, especially in automating equipment related to microscopy[8-10]. In this study, we have proposed a method to analyze hexagonal structures in optical microscope images. The process involves detecting individual hexagons, extraction of the regions consisting of clusters of hexagons, and the validation of the images acquired by employing the Haar cascade classifier[11], K-means clustering algorithm[12], and a Convolutional Neural Network(CNN)[13], respectively. An overview of the workflow is outlined in figure S13.

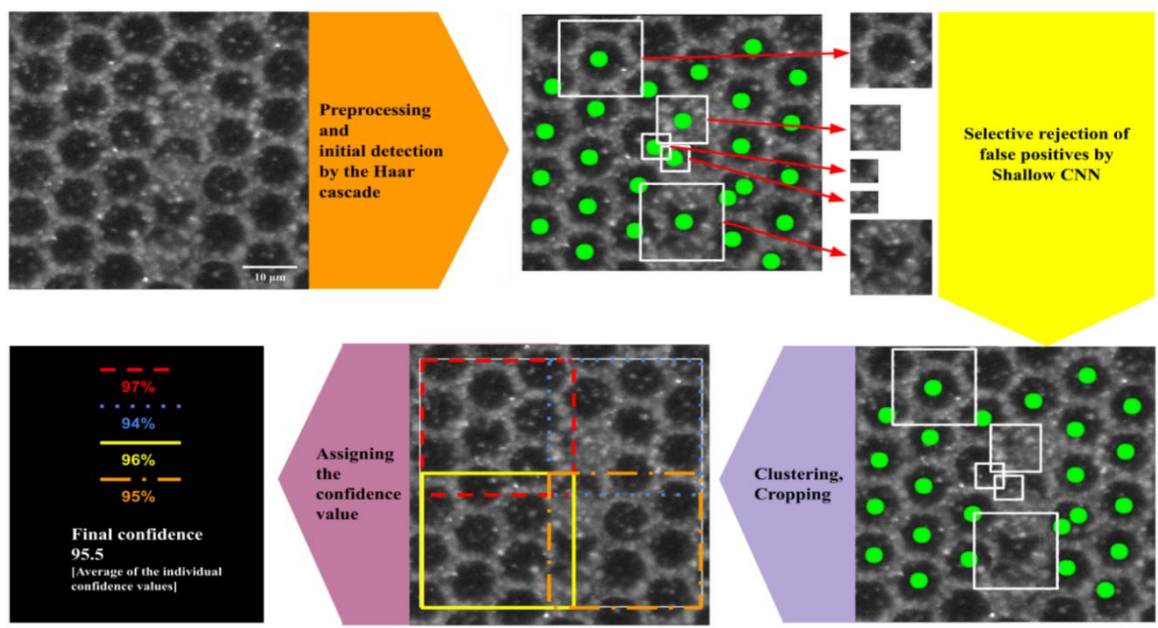

Figure S13: The workflow of the pattern recognition protocol.

## S2. The workflow

The hexagon detecting workflow consists of five algorithms:

**S2.1. Algorithm 1:** Data preparation.

The images obtained from the optical microscope are processed before the Haar cascade stage (Algorithm 2 Figure S14b). The scale on the images is removed before augmentation. Augmentation is required as it helps to increase the total number of images that can be used for training[14] the Haar cascade; it is done by rotating the images and by inducing intensity changes (dimming and brightening the images). The data preparation process is shown in Algorithm 1 Figure S14 (a).

**S2.2. Algorithm 2:** Haar Cascade Training and selective rejection by the shallow CNN.

The Haar cascade classifier is a supervised method in which multi-resolution sub-windows over a region in an image are sequentially subjected to a decision-making process at a series of nodes in the cascade[11]. A rejection decision terminates the detection process, while a temporary acceptance allows the detection process to continue. Since the instances where a non-hexagonal structure is present in the detection window is more compared to a hexagonal structure being present in the detection window, the algorithm is computationally efficient. A positive set and a negative set, each consisting of 500 images, are created. The negative set consists of images representing the background and everything other than the object of Ainterest (Undistorted hexagon). The positive set consists of the negative set images superimposed with the object of interest (Figure S14 (b)). The cascade.xml file created via the training process contains the information required to detect hexagonal structures. The hexagons are detected using the cascade.xml file; the x and y coordinates of the top left corner and the width and height are obtained via the detection process. The cropped-out individual hexagons are resized into 28 x 28 images before being subjected to validation via a shallow CNN. The shallow CNN will selectively accept those detections which can be assigned confidence higher than 0.8 (since it is a reasonable threshold). The centres of the accepted detections are calculated via the equations (S1) and (S2).

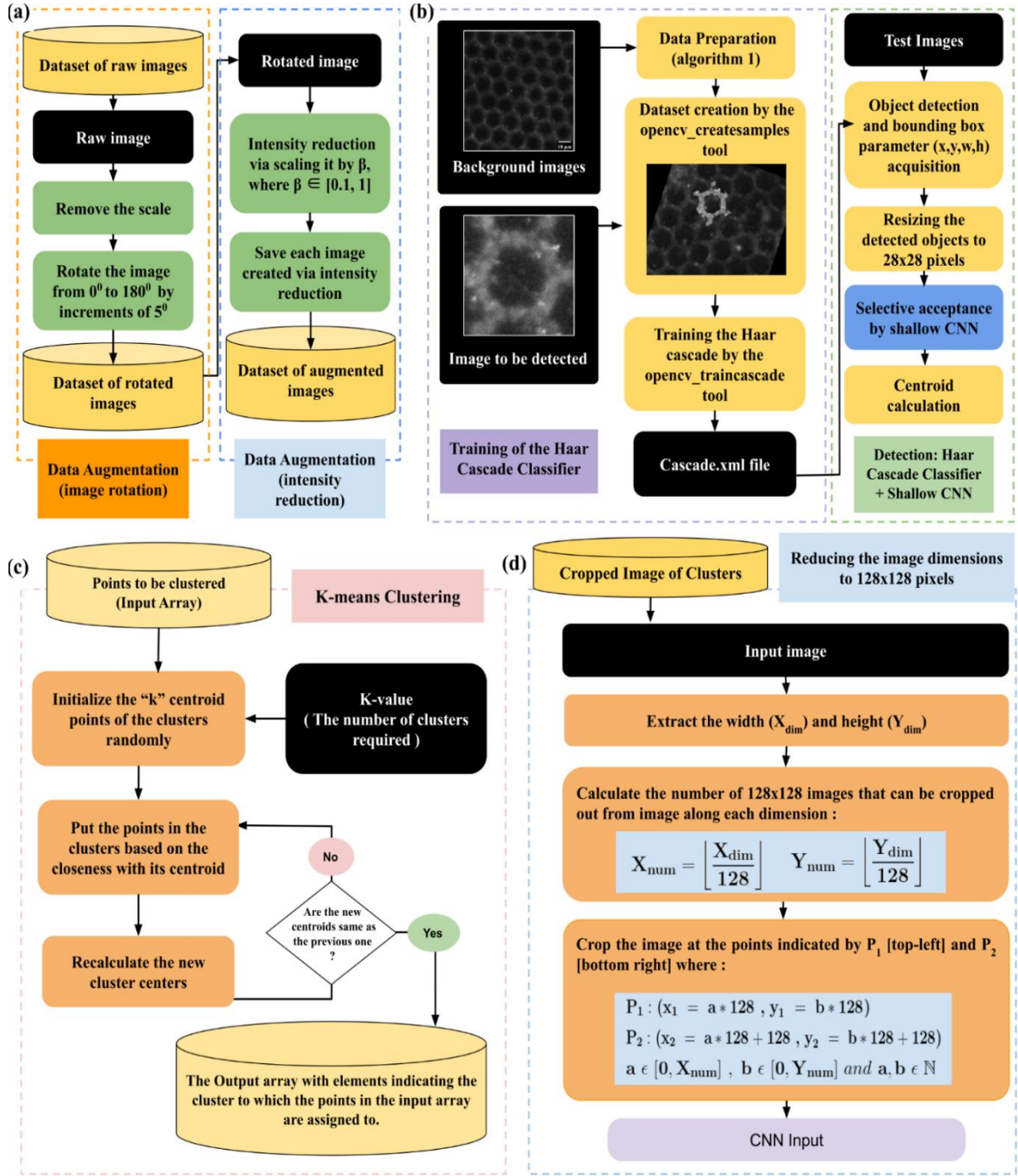

Figure S14: (a) Algorithm 1: Data preparation. (b) Algorithm 2: Haar Cascade Training (c) Algorithm 3: K-means Clustering (d) Algorithm 4: Creating the input for the CNN

$$X_{center} = X_{BoundingBox} + \frac{width}{2} \qquad (S1)$$

$$Y_{center} = Y_{BoundingBox} + \frac{height}{2} \qquad (S2)$$

**S2.3. Algorithm 3:** K-means Clustering

The clustering of position coordinates is done via the k-means clustering algorithm, an unsupervised machine learning algorithm[12]. The position coordinates are categorized into the specified "k" number of clusters. The categorization is done by minimizing the variance of the position coordinates (within a cluster) with respect to the coordinates of the cluster's centroid. The procedure is outlined in Algorithm 3 (Figure S14 (c)). The coordinates of points in each cluster can be used to crop out the corresponding cluster of hexagonal structures from the whole image. The cropped cluster is rejected if it has less than 5 hexagons.

**S2.4. Algorithm 4:** Creating the input for the Deep-CNN

The cropped-out clusters of hexagons should be converted into a set of images with a dimension of 128x128 in pixels in order to match the input of the CNN. The procedure for the same is mentioned in Algorithm 4 (Figure S14 (d)).

**S2.5. Algorithm 5:** Assigning the confidence value via the CNN

**S2.5.1. Deep-CNN architecture:** The Convolutional Neural Network (CNN) architecture consists of Convolutional and Max-pooling layers, followed by a series of fully connected layers[13]. The main components of the convolutional layer are the input, kernel, and feature map matrices. The feature map is the input of the subsequent layer and is created by sliding the kernel over the input matrix with a certain stride (the convolution operation)[15]. The convolution operation identifies the same features in different regions of the image[16]. In the convolutional layers, the convolutional operation is followed by max-pooling operations; it is a non-linear down-sampling operation in which a kernel via convolution extracts high activation regions from the input[17]. Due to max-pooling, translational invariance is induced on the features of the input image. Dropout-regularization was applied after the max-pooling operation. The dropout regularization involves the activation of only a certain subset of nodes in a layer to reduce the chance of the network learning redundant features[18]. Fully connected layers are introduced in the CNN architecture following the convolutional layers. In fully connected layers, every node in one layer is connected to all the nodes in the successive layer. This helps in reducing the number of nodes in a controlled manner[19]. Dropout regularization was also applied to the fully connected layers. The CNN architecture used in our study consists of two convolutional layers. The kernel matrices used in the two layers for the convolution operation are 16 and 32 in number, respectively, with a kernel dimension of 5 x 5 and a stride of 2. The max-pooling operation utilized 2 x 2 kernels with a stride of 2 in both layers. The Rectified Linear Unit (ReLU) activation function was used between the layers[20]. For an input vector I, weight matrix α and bias b, it has the mathematical form as given by equation (S3).

$$ReLU = \max(0, \alpha^T I) \qquad (S3)$$

A truncated normal distribution with zero mean and a standard deviation of 0.1 was used to initialize the weights. The biases were initialized with a constant value of 0.1. With a learning rate of $10^{-4}$, Adam Optimizer was used for weight and bias optimization during training. Adam Optimizer has high computational efficiency and requires less memory, which leads to its selection for our study[21]. The cross-entropy error function[22] was used since we had labels that followed the one-hot encoding. The output (prediction) was calculated using a SoftMax function[23] with the mathematical form as given in equation (S4).

$$Softmax_m = \frac{e^{\alpha_m^T I}}{\sum_{k=1}^{n} e^{\alpha_k^T I}} \qquad (S4)$$

Where $\alpha^T_m$ is the weight of the $m^{th}$ node, I is the input, and n is the number of output nodes.

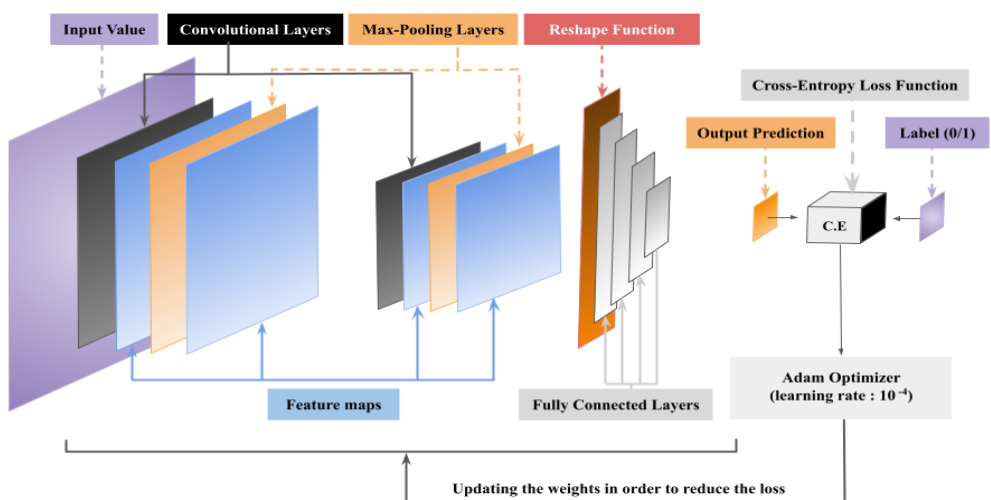

Figure S15: Schematic of the deep CNN architecture.

The CNN architecture is represented in Figure S15. The Deep-CNN was trained on 300 images (150 positive and 150 negative images). The validation and testing of the Deep-CNN was done using validation and test sets, with each of them containing of 150 images (75 positive and 75 negative images). The Accuracy-Epoch plot and the Confusion matrix[24] were used to analyze the performance of the CNN[16]. The metrics used in the confusion matrix involved the true-positive (TP), false-positive (FP), true-negative (TN) and false-negative (FN) values. In binary classification, the samples are labelled as either positive (1) or negative (0). If the prediction and target value are both negative, it is TN; if the prediction and target value are both positive, it is TP. If the prediction value is positive while the target value is negative, its FP; if the prediction value is negative, and the target value is positive, its FN. The accuracy after each epoch was calculated on the validation data set. After the training process involving 50 epochs, the model's average performance was indicated using the Confusion Matrix calculated using predictions made by the Deep-CNN on the validation set and the test set. Additional performance metrics derived from the TP, TN, FN, and FP values evaluated using equations (S5) - (S14).

$$Recall\ (True\ Positive\ Rate) = \frac{TP}{TP + FN} \qquad (S5)$$

$$Negative\ Predictive\ Value = \frac{TN}{TN + FN} \qquad (S11)$$

$$Fall\ Out\ (False\ Positive\ Rate) = \frac{FP}{FP + TN} \qquad (S6)$$

$$False\ Omission\ Rate = \frac{FN}{FN + TN} \qquad (S10)$$

$$Specificity\ (True\ Negative\ Rate) = \frac{TN}{TN + FP} \qquad (S7)$$

$$Threat\ Score = \frac{TP}{TP + FN + FP} \qquad (S12)$$

$$False\ Discovery\ Rate = \frac{FP}{FP + TP} \qquad (S8)$$

$$Miss\ Rate\ (False\ Negative\ Rate) = \frac{FN}{TP + FN} \qquad (S13)$$

$$Precision\ (Positive\ Predictive\ Value) = \frac{TP}{TP + FP} \qquad (S9)$$

$$F1\ Score = \frac{2TP}{2TP + FP + FN} \qquad (S14)$$

$$Accruacy = \frac{TP + TN}{TP + TN + FP + FN} \qquad (S15)$$

## S2.5.2. Assigning the confidence value.

The "confidence value" refers to the scalar output obtained from the trained Deep-CNN (where the weights are fixed via the training process) when an image of the preferred dimension is provided as an input. The overall confidence ($C_{overall}$) is calculated using Equation (S16).

$$C_{overall} = \frac{1}{n}\sum_{i=1}^{n} C_i \qquad (S16)$$

Where $C_i$ is the scalar output obtained from the Deep-CNN when images of dimension 128 x 128 (obtained as the output of Algorithm 4 Figure S14 (d)) are provided as input, n is the number of input images.

## S3. Computational Requirement

The Haar Cascade was trained using a quad core i3-5005U CPU with 4 gigabytes of RAM. The CNN was trained using Google Colab, an online GPU lending service. The algorithm was implemented in the python programming language. The CNN architectures were created using TensorFlow Version 1.

## S4. Results and discussion

### S4.1 Performance of the Haar cascade

The Haar cascade was tested on a set of 62 images, out of which 2 were desirable (positive), and 60 were undesirable (negative). The number of positives being substantially smaller than the number of negatives is attributed to the difficulty in acquiring the images. The Haar cascade performed well in detecting the presence of hexagonal structures in the positive images (Figure S16(d)). Ideally, the number of hexagonal structures detected in the negative images should be zero. However, the Haar Cascade being a boosting technique is prone to detecting False positives.

| Performance Metric | Our model on the validation data set | Our model on the test data set | Ideal Value |
|---|---|---|---|
| Recall | 0.9867 | 1 | 1 |
| Specificity | 0.9067 | 0.9867 | 1 |
| Precision | 0.9136 | 0.9868 | 1 |
| Negative Predictive Value | 0.9855 | 1 | 1 |
| Miss Rate | 0.0133 | 0 | 0 |
| Fall Out | 0.0933 | 0.0133 | 0 |
| False Discovery Rate | 0.0864 | 0.0132 | 0 |
| False Omission Rate | 0.0145 | 0 | 0 |
| Threat Score | 0.9024 | 0.9868 | 1 |
| F1 Score | 0.9487 | 0.9934 | 1 |

Table S4: The performance metrics calculated from the TP, TN, FP and FN values using equations (S5) – (S14)

The detection in the negative images was classified into three categories:

• **Zero Detections:** In these negative images, the Haar cascade detected zero hexagonal structures (Figure S16(e)). This category included 22 images from the total 60 images used.

• **Rejectable Detections:** In these negative images, the Haar cascade detected hexagonal structures, but the number of hexagons in a cluster was less than or equal to 5 (Figure S16(f)). These images could be rejected during the clustering stage.

• **Non-Rejectable Detections:** The Haar cascade detected hexagonal structures, and the number of hexagons in a cluster was more than 5 (Figure S16(g)). These images cannot be rejected during clustering since the number of hexagons detected is close to the number of hexagons detected in the positive images. These images can be rejected using the shallow CNN.

When individual samples are considered, the percentage of true positive detections (PTP) ranged from 55% to 100% (Figure S16 (a)-(c))

**S4.2. Shallow CNN architecture, training and performance**

**Architecture:** The shallow CNN architecture consists of two convolutional layers (with 8 and 16 convolutional filters having a dimension of 3x3 and a stride of 1) and one fully connected layer. The activation used is a ReLU function (Equation S3) with a SoftMax function as the final output node (Equation (S4)).

**Training:** The images returned by the Haar Cascade are of different sizes (Figure S16(h), S16(i) and S16(j)); hence they are resized into images of 28 x 28 pixel dimensions (Figure S16(h), S16(i) and S16(j)). The resized images of complete hexagons were considered as the positive set. The resized images of background noise, corners, and edges were taken as the negative set. The positives and negatives were augmented to create 600 images which formed the data set for training and testing the shallow CNN. The unbiased training and testing sets created from this data set contained 300 images each.

**Performance:** The Accuracy vs. Epoch plot was used to assess the Performance of the shallow CNN (Figure S16(k)). The average accuracy on the validation data set is approximately 95 %.

**S4.3. Performance of the Deep Convolutional Neural Network (Deep-CNN)**

• **Performance on the Validation data set:** The performance of the Convolutional Neural Network was evaluated using the Accuracy vs Epoch plot and the Confusion Matrix. The Accuracy (Figure S16(l)) and Confusion Matrix (Figure S16(m)) were calculated using the Validation set consisting of 150 images. Out of these 150 images, 75 images were positives, and 75 were negative. On average, 68 of the 75 negative images were detected correctly (TN), and 74 of the 75 positive images were detected correctly (TP). The performance metrics calculated from the TP, TN, FP, and FN values are given in Table S4.

• **Performance on the Test data set:** The performance on the Test data set was evaluated after the model was trained for 50 epochs. The confusion matrix for the CNN is given in figure S16(n). The performance metrics derived from the Confusion matrix as in the validation case are given in Table S4.

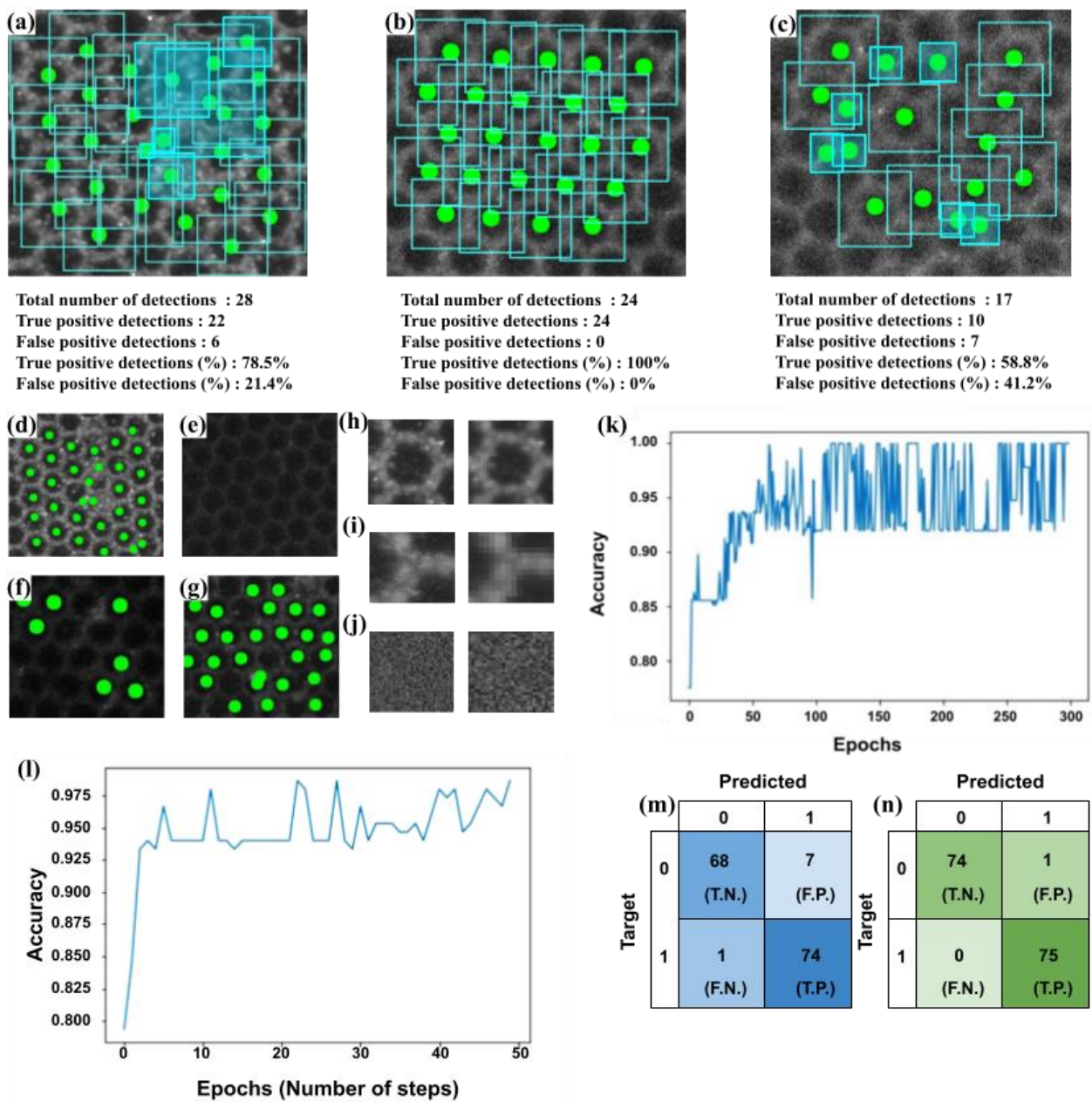

Figure S16: The green dots represent the centres of the bounding box drawn around the hexagons detected via Haar Cascade. (a)-(c) The detections from the Haar Cascade indicating the percentage of true positives (PTP) and percentage of false positives (PFP) [indicated with the transparent blue pane]. (a) The PTP : 78.5% and PFP : 21.4% , (b) The PTP : 100% and PFP : 0%, (c) The PTP : 58.8% and PFP : 41.2%. (d) Detection of hexagons in the positive image., (e) Zero detection case., (f) Rejectable detection case., (g) Non-Rejectable detection., (h) The complete hexagon returned by the Haar cascade and its resized form., (i) Fragments of hexagons detected by the Haar cascade and its resized form., (j) Noisy patch returned by the Haar cascade and its resized form., (k) Accuracy vs Epoch plot for the shallow CNN., (l) Accuracy vs Epoch plot of the deep-CNN., (m) Confusion matrix generated by the deep-CNN on the Validation data set., (n) Confusion matrix generated by the deep-CNN on the Test data set.

## S5. Conclusion

We have developed a workflow using cascade, clustering and neural net algorithms to identify and analyze the quality of hexagonal objects in optical microscope images. The occurrence of False positives in the Haar Cascade detections has been compensated by utilizing CNNs. The performance of workflow is at par with the ideal scenario as provided in Table S4. Since it performs well with regard to detecting hexagonal structures, the algorithm is trained on images of other objects (nano-particles, nano-prisms etc.), can be extended to detect the same. The workflow can be utilized in order to create detectors for a certain object, and the computational requirements for the same are minimal. Hence the utility of such a workflow is unparalleled in scenarios where object detection and analysis is the main theme.